\documentclass[journal]{IEEEtran}
\usepackage{cite}
\usepackage{amsmath,amssymb,amsfonts}
\usepackage{algorithmic}
\usepackage{graphicx}
\usepackage{textcomp}

\usepackage{xcolor}
\usepackage{times}
\usepackage{epsfig}
\usepackage{algorithm}
\usepackage[nice]{nicefrac}
\usepackage{multirow}

\begin{document}
\title{Multi-level Context Gating of Embedded Collective Knowledge for Medical Image Segmentation}

\author{Maryam~Asadi-Aghbolaghi, Reza~Azad, Mahmood~Fathy, and~Sergio~Escalera
\thanks{The first two authors contributed equally.}
\thanks{This work is partially supported by the Spanish project TIN2016-74946-P (MINECO/FEDER, UE) and CERCA Programme/Generalitat de Catalunya. We gratefully acknowledge the support of NVIDIA Corporation with the donation of the GPU used for this research. This work is partially supported by ICREA under the ICREA Academia programme.}
\thanks{M. Asadi-Aghbolaghi, and M. Fathy are with the School of Computer Science, Institute for Research in Fundamental Sciences (IPM), Tehran, Iran, (e-mail:masadi@ipm.ir, mahfathy@ipm.ir).} 
\thanks{R. Azad is with the Department of Computer Engineering, Sharif University of Technology, Tehran, Iran, (e-mail: rezazad68@gmail.com).}
\thanks{S. Escalera is with the Universitat de Barcelona and Computer Vision Center, Barcelona, Spain. (email: sergio@maia.ub.es)}}

\maketitle

\begin{abstract}
Medical image segmentation has been very challenging due to the large variation of anatomy across different cases. Recent advances in deep learning frameworks have exhibited faster and more accurate performance in image segmentation. Among the existing networks, U-Net has been successfully applied on medical image segmentation.
In this paper, we propose an extension of U-Net for medical image segmentation, in which we take full advantages of U-Net, Squeeze and Excitation (SE) block, bi-directional ConvLSTM (BConvLSTM), and the mechanism of dense convolutions. 
(I) We improve the segmentation performance by utilizing SE modules within the U-Net, with a minor effect on model complexity. %These blocks adaptively recalibrate the channel-wise feature responses by modeling the explicit self-mapped interdependencies between channels.
These blocks adaptively recalibrate the channel-wise feature responses by utilizing a self-gating mechanism of the global information embedding of the feature maps. 
(II) To strengthen feature propagation and encourage feature reuse, we use densely connected convolutions in the last convolutional layer of the encoding path. % to collect all the knowledge learned in the encoding path. 
(III) Instead of a simple concatenation in the skip connection of U-Net, we employ BConvLSTM in all levels of the network to combine the feature maps extracted from the corresponding encoding path and the previous decoding up-convolutional layer in a non-linear way. 
%The proposed model is evaluated on six datasets of: retinal blood vessel segmentation, 2017 and 2018 versions of skin lesion segmentation, lung nodule segmentation, $PH^2$, and Data Science Bowl 2018 achieving state-of-the-art performance.
The proposed model is evaluated on six datasets DRIVE, ISIC 2017 and 2018, lung segmentation, $PH^2$, and cell nuclei segmentation, achieving state-of-the-art performance.

\end{abstract}


\section{Introduction}
\label{sec:introduction}

\begin{figure}
\centering
%\vspace{-5mm}
\includegraphics[width=0.45 \textwidth]{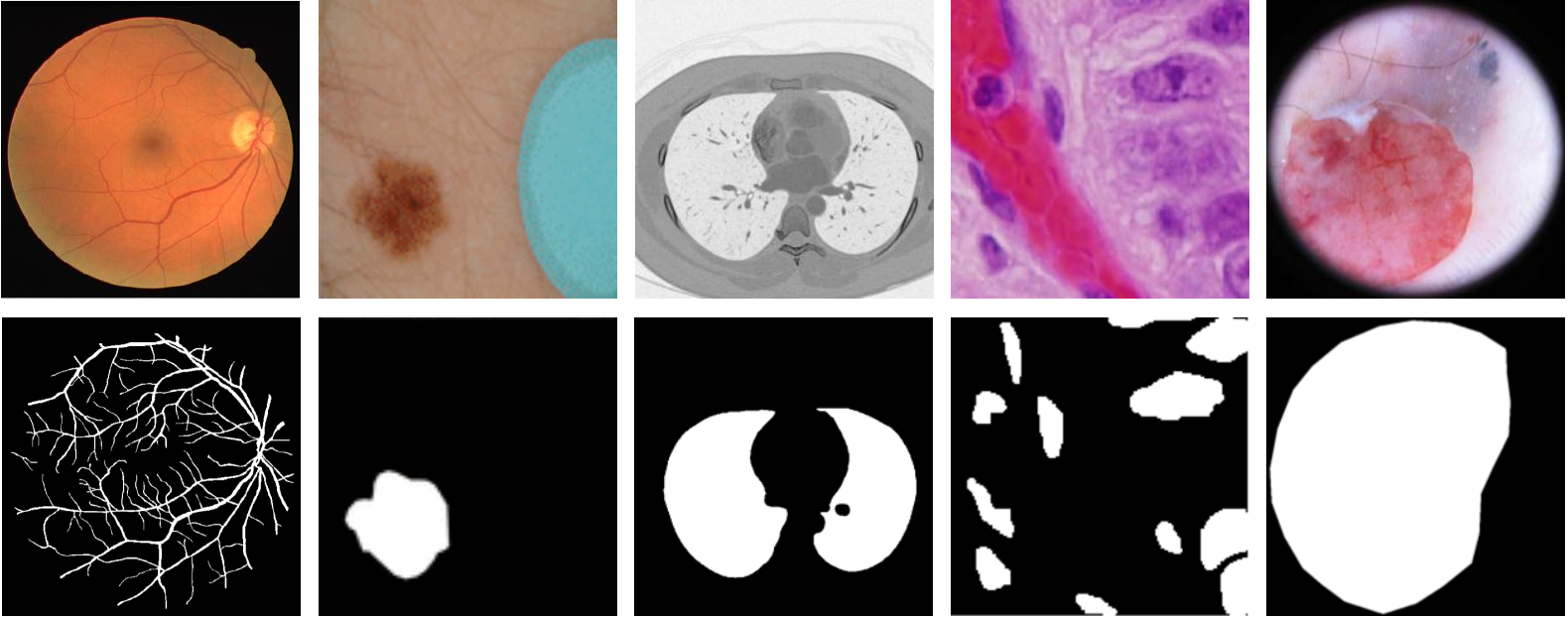}
%\vspace{-4mm}
\caption{Different applications of medical image segmentation.} \label{fig:app}
\vspace*{-\baselineskip}
\end{figure}

\IEEEPARstart{M}{edical} images play a key role in medical treatment and diagnosis. 
The goal of Computer-Aided Diagnosis (CAD) systems is providing doctors with more precise interpretation of medical images to follow-up of many diseases and have better treatment of a large number of patients. %The medical imaging includes different kinds of imaging techniques like Magnetic Resonance Imaging (MRI), ultrasound, and Computer Tomography (CT).
Moreover, accurate and reliable processing of medical images results in reducing the time, cost, and error of human-based processing. A critical step in numerous medical imaging studies is image segmentation. Medical image segmentation is the process of partitioning an image into multiple meaningful regions. Due to the complex geometry and inherent noise value of medical images, segmentation of these images is difficult. 
Interest in medical image segmentation has grown considerably in the last few years. This is due in part to the large number of application domains, like segmentation of blood vessel, skin cancer, lung, and cell nuclei (Figure \ref{fig:app}). %, for this type of technology.

For instance, segmentation of blood vessels %(Figure \ref{fig:app_Ret})
will help to detect and treat many diseases that influence the blood vessels. Width and curves of retinal blood vessel show some symptoms about many diseases. %, which can be useful for diagnosis. 
Early diagnosis of many sight-threatening diseases is vital since lots of these diseases like glaucoma, hypertension and diabetic retinopathy cause blindness among working age people. 
%Skin lesion segmentation has a critical role in the early and accurate diagnosis of skin cancer by computerized systems. %Skin cancer is one of the most common cancer types. % in over the world.
Skin lesion segmentation helps to detect and diagnosis the skin cancer in the early stage. One of the most deadly form of skin cancer is melanoma, which is the result of unusual growth of melanocytes. Dermoscopy, captured by the light magnifying device and immersion fluid, is a non-invasive imaging technique providing with a visualization of the skin surface. The detection of melanoma in dermoscopic images by the dermatologists may be inaccurate or subjective. %On the other hand, Early diagnosis of the melanoma is thus critically important in terms of treatment. 
If melanoma is detected in its early stages, the five-year relative survival rate is $92\%$ \cite{siegel2018jemal}. %As a result, skin lesion segmentation (Figure \ref{fig:app_ISIC}) has a critical role in the early and accurate diagnosis of skin cancer by computerized systems.

% \begin{figure}
% \centering
% %\vspace{-5mm}
% \includegraphics[width=0.5 \textwidth]{Data.pdf}
% %\vspace{-4mm}
% \caption{Different applications of medical image segmentation like blood vessel segmentation, skin cancer segmentation, and lung segmentation..} \label{fig:applications}
% \vspace*{-\baselineskip}
% \end{figure}

% \begin{figure}
% \centering
% %\vspace{-5mm}
% \includegraphics[width=0.5 \textwidth]{Retina_Intro.pdf}
% %\vspace{-4mm}
% \caption{An application of medical image segmentation for blood vessel segmentation.} \label{fig:app_Ret}
% \vspace*{-\baselineskip}
% \end{figure}

% \begin{figure}
% \centering
% %\vspace{-5mm}
% \includegraphics[width=0.5 \textwidth]{ISIC_Intro.pdf}
% %\vspace{-4mm}
% \caption{An application of medical image segmentation for skin cancer segmentation.} \label{fig:app_ISIC}
% \vspace*{-\baselineskip}
% \end{figure}

% \begin{figure}
% \centering
% %\vspace{-5mm}
% \includegraphics[width=0.5 \textwidth]{Lung_Intro.pdf}
% %\vspace{-4mm}
% \caption{An application of medical image segmentation for lung segmentation..} \label{fig:app_Lung}
% \vspace*{-\baselineskip}
% \end{figure}

The first vital step of pulmonary image analysis is identifying the boundaries of lung from surrounding thoracic tissue on CT images, called lung segmentation. %(Figure \ref{fig:app_Lung}) 
%which is very crucial for identifying lung related diseases. 
It can also be applied to lung cancer segmentation. 
Another application of medical image segmentation is cell nuclei segmentation. All known biological lives include a fundamental unit called cell. %Since all of the human body’s cells contain a nucleus full of DNA, identifying the cells’ nuclei is a crucial task for most medical image analyses. 
By segmentation of nuclei in different situations, we can understand the role and function of the nucleus and the DNA contained in cell in various treatments.

%Like other fields of research in computer vision, 
Deep learning networks achieve outstanding results and use to outperform non-deep state-of-the-art methods in medical imaging. 
%Deep neural networks are mostly utilized in classification tasks, where the output of the network is a single label or probability values associated labels to a given input image. %AlexNet, VGG, Residual Net, and DenseNet are among the networks for classification task. 
%These networks work fine thanks to some structural features \cite{alom2018} such as: activation function, different efficient optimization algorithms, and dropout as a regularizer for the network. 
These networks require a large amount of data to train and provide a good generalization behavior given the huge number of network parameters. A critical issue in medical image segmentation is the unavailability of large (and annotated) datasets. In medical image segmentation, per pixel labeling is required instead of image level label. 
%%%%To mitigate this problem, many local regions of the input image are selected and utilized as the input data to networks.
Fully convolutional neural network (FCN) \cite{long2015fully} was one of the first deep networks applied to image segmentation.

Ronneberger et al. \cite{ronneberger2015} extended this architecture to U-Net, achieving good segmentation results leveraging the need of a large amount of training data. Their network consists of encoding and decoding paths. In the encoding path a large number of feature maps with reduced dimensionality are extracted. % from the input data.
The decoding path is used to produce segmentation maps (with the same size as the input) by performing up-convolutions.
 %The most important modification is mainly about the skipping connections. In this part of the network, extracted feature maps from the previous up-convolutional layer are concatenated with the extracted features from the corresponding encoding path.
Many extensions of U-Net have been proposed so far \cite{alom2018,oktay2018,azad2019bi}.
In some of them, the extracted feature maps in the skip connection are first fed to a processing step (e.g. attention gates \cite{oktay2018}) and then concatenated. The main drawback of these networks is that the processing step is performed individually for the two sets of feature maps, and these features are then simply concatenated. 

%In this paper, we aim at proposing a deep learning approach in order to segmenting medical images as an expert would do.

In this paper,  we propose %Squeeze and Excitation U-Net with Densely connected convolutions (\textit{MCGU-Net}),
\textit{Multi-level Context Gating U-Net} (MCGU-Net)
 an extended version of the U-Net, by including BConvLSTM \cite{song2018pyramid} in the skip connection, using SE mechanism in the decoding path, and reusing feature maps with densely convolutions. A VGG backbone is employed in the encoding path to make it possible to use pre-trained weights on large datasets. The feature maps from the corresponding encoding layer have higher resolution while the feature maps extracted from the previous up-convolutional layer contain more semantic information. Instead of a simple concatenation, combining these two kinds of features with non-linear functions in all levels of the network may result in more precise segmentation. Therefore, in this paper we extend the U-Net architecture by adding multi-level BConvLSTM in the skip connection. % to combine these two kinds of feature maps.

Inspired by the effectiveness of the recently proposed squeeze and excitation modules \cite{hu2018squeeze} on image classification, we modify the U-Net by inserting these blocks in the decoding path. SE modules allow the network to  recalibrate the feature map to have more attention on useful channels by assigning different weights to various channels of feature maps based on to their relationship by employing a context gating mechanism. % and interdependencies. 
By using global embedding information, these modules help the network to boost informative and meaningful features, while suppressing weak ones. 
Having a sequence of convolutional layers may help the network to learn more kinds of features; however, in many cases, the network learns redundant features. To mitigate this problem and enhance information flow through the network, we utilize the idea of densely connected convolutions \cite{huang2017densely}. In the last layer of the contracting path, convolutional blocks are connected to all subsequent blocks in that layer via channel-wise concatenation. %Features which are learned in each block are passed forward to the next block. 
This strategy helps the method to learn a diverse set of features based on the “collective knowledge” gained by previous layers. %, and therefore, avoiding learning redundant features. 
Furthermore, we accelerate the convergence speed of the network by employing BN after the up-convolution filters.

%Feature maps with the highest semantic information are produced in the last step of the encoding layer. It is well-known that convolutions just capture local information. In order to capture more global information, we augment the last convolutional operators in the encoding path with a self-attention mechanism, as proposed by Bello et al. \cite{bello2019attention}. To do that, the convolutional feature maps are concatenated with a set of feature maps produced via self-attention. 

We evaluate the proposed MCGU-Net on four different applications retinal blood vessel segmentation (DRIVE dataset), Skin lesion segmentation (three datasets of $PH^2$, ISIC 2017 and 2018), lung nodule segmentation (Lung dataset), and cell nuclei segmentation (Data Science Bowl 2018). The experimental results demonstrate that the proposed network achieves superior performance than state-of-the-art alternatives. \footnote{Source code is available on https://github.com/rezazad68/BCDU-Net.}

\section{Related work}
%One of the most crucial tasks in medical imaging is semantic segmentation. Before the revolution of deep learning in computer vision, traditional handcrafted features were exploited for semantic segmentation. 
During the last few years, deep learning-based approaches have outstandingly improved the performance of classical image segmentation strategies. % and recognition algorithms in different tasks of computer vision. 
%It is not very difficult to transfer these advances to medical images segmentation. 
Based on the exploited deep architecture, we divide these approaches into three groups. %, CNN, FCN, and RNN.

\subsection{Convolutional Neural Network (CNN)}
Cui et al. \cite{cui2016} exploited CNN for automatic segmentation of brain MRI images. The authors first divided the input images into some patches and then utilized these patches for training CNN. To handle an arbitrary number of modalities as the input data, Kleesiek et al. \cite{kleesiek2016} proposed a 3D CNN for brain lesion segmentation. To process MRI data, the network consists of four channels: non-enhanced and contrast-enhanced T1w, T2w and FLAIR contrasts. Roth et al. \cite{roth2015deeporgan} proposed a multi-level deep convolutional networks for pancreas segmentation in abdominal CT scans as a probabilistic bottom-up approach.

\subsection{Fully Convolutional Network (FCN)}
A problem of the CNN models for segmentation is that the spatial information of the image is lost when the convolutional features are fed into the fc layers. To overcome this problem the FCN was proposed \cite{long2015fully}. This network is trained end-to-end and pixels-to-pixels, in which all fc layers of the CNN architecture are replaced with convolutional and deconvolutional to keep the original spatial resolutions. %Therefore, the original spatial dimension of the features maps are recovered while the network is performing the segmentation task.
%FCN has been frequently utilized for segmentation of medical and biomedical images \cite{zhou2016three,zhou2017fixed}. 
Zhou et al. \cite{zhou2016three} exploited FCN for segmentation of anatomical structures on 3D CT images. An FCN with convolution and de-convolution parts is trained end-to-end, performing voxel-wise multiple-class classification to map each voxel in a CT image to an anatomical label. Drozdzal et al. \cite{drozdzal2016} proposed very deep FCN by using short skip connections. The authors showed that a very deep FCN with both long and short skip connections achieved better result than the original one.
Roth et al. \cite{roth2018application} proposed to employ 3D FCN in a cascaded fashion for segmentation of the organs and vessels in CT images.
%In that method the segmentation of the anatomical structures (including multiple organs) in a CT image (generally in 3D) is performed by majority voting of the semantic segmentation of multiple 2D slices drawn from different viewpoints. Roth et al. \cite{roth2018application} proposed to employ 3D FCN in a cascaded fashion for segmentation of the organs and vessels in CT images. The authors utilized a two-stage, coarse-to-fine architecture of FCN. In the first stage, about $40\%$ of the voxels (automatically generated mask of the patient’s body) are processed by FCN. This amount of voxels is reduced to $10\%$, and the network is allowed on more detailed segmentation. In original FCN, long skip connections are utilized to skip features from the contracting path to the expanding one.

U-Net, %proposed by Ronneberger et al. 
\cite{ronneberger2015}, is one of the most popular FCNs for medical image segmentation. %This network consists of contracting and expanding paths. 
%In the contracting path, convolution filters followed by ReLU and $2\times2$ max-pooling operators are applied on the input data. In the contracting path, convolution, ReLU and max-pooling operators are applied on the input data. In the expanding path, de-convolution are utilized to up-sample the feature maps. The feature maps were copied and cropped from encoding units to the decoding ones.
It has some advantages than the other segmentation-based networks \cite{alom2018}. It works well with few training samples and the network is able to utilize the global location and context information at the same time.
%Different extension versions of U-Net have been proposed for segmentation task. 
Milletari et al. \cite{milletari2016} proposed V-Net, a 3D extension version of U-Net to predict segmentation of a given volume at once. V-Net is an end-to-end 3D image segmentation network based on a volumetric (MRI volumes). %, fully convolutional, neural network. 
3D U-Net \cite{cciccek20163d} is proposed for processing 3D volumes instead of 2D images. In which, all 2D operations of U-Net are replaced with their 3D counterparts. 
%In that network, all 2D operations in U-Net are replaced with their 3D counterparts.%, i.e., 3D convolutions, 3D max pooling, and 3D up-convolutional layers. 
In \cite{kayalibay2017}, %two modifications have been applied on the standard U-Net. 
the authors combine multiple segmentation maps that are created at different scales. Moreover, to forward feature maps from one stage of the network to the other one, element-wise summation is utilized.
A dual pathway 3D CNN (with 11 layers) \cite{kamnitsas2017} was proposed for brain lesion segmentation in multi-modal brain MRI. In this model, input images at multiple scales are fed simultaneously to a FCN. 
%Moreover, the adjacent image patches are processed into one pass through the network. The authors also employed a 3D fully connected Conditional Random Field which effectively removes false positives as a post-processing step.
Li et al. proposed High-Res3DNet \cite{li2017compactness}, which is a high-resolution, compact convolutional network for volumetric image segmentation. % which include high spatial resolution feature maps throughout the layers.

%VoxResNet \cite{chen2016voxresnet}, a deep voxel-wise residual network, was proposed for brain segmentation from MR. This 3D network is inspired by deep residual learning, performing summation of feature maps from different layers. 

\subsection{Recurrent Neural Network (RNN)}
One of the most used neural networks for processing a sequence is RNN, which can take into account the temporal data using recurrent connections in hidden layers. It has been successfully applied for modeling short- and long-temporal sequences. These networks are able to model the global contexts and improve semantic segmentation.Different RNN based deep network have been proposed for semantic segmentation. 
Pinheiro et al. \cite{pinheiro2014} proposed %an end-to-end feed forward 
a deep network consisting of an RNN that can take into account long range label dependencies in the scenes while limiting the capacity of the model. Visin et al. \cite{visin2016reseg} proposed ReSeg for semantic segmentation. In that network, the input images are processed with a pre-trained VGG-16 model and its resulting feature maps are then fed into one or more ReNet layers. %Finally an up-sampling layer is employed for resizing the feature map.
DeepLab architecture \cite{chen2017deeplab} contains a deep convolutional neural network in which all fully connected layers are replaced by convolutional layers and then the feature resolution is increased through atrous convolutional layers.
Alom et al. \cite{alom2018} proposed Recurrent Convolutional Neural Network (RCNN) and Recurrent Residual Convolutional Neural Network (R2CNN) based on U-Net models for medical image segmentation. %The residual unit helps the network in training. The feature accumulation with recurrent residual convolutional layers improve the feature representation for segmentation tasks.
%Bai et al. \cite{bai2018recurrent} combined an FCN with an RNN for medical image sequence segmentation, which is able to incorporate both spatial and temporal information for MR images.
Gao \cite{gao2018fully} proposed an end to end combination of FCN and RNN with long short-term memory (LSTM) units for 4D segmentation of MRI images.

He et al. \cite{hu2018squeeze} introduced the Squeeze and Excitation (SE) network for image classification which models the explicit relationship between the channels of a feature map. % its convolutional features. %The SE block can be integrated in any CNN model. 
In these modules, the convolutional features are first passed through a squeeze operation in which global average pooling is exploited to produce channel descriptor. The output of the aggregation is then fed to an excitation operation to generate a set of per-channel modulation weights. These weights are utilized to recalibrate the feature map to emphasize on useful channels. %The SE strategy have been employed in many CNN networks \cite{roy2018concurrent,li2018recurrent}.

%SE modules have been also utilized in U-Net. Rundo et al. \cite{rundo2019use} employed this block in the skip connection to before the concatenation of feature maps from two encoding and decoding path. In \cite{zhu2018anatomynet} the authors, replaced the standard convolutional layers in U-Net with SE blocks.

In this paper, MCGU-Net is proposed as an extension of U-Net, showing better performance than state-of-the-art alternatives. % for the segmentation task. We take advantages of BConvLSTM, SE modules, and dense convolutions to improve the representational power of the original U-Net. 
The BConvLSTM is employed in the skip connection to combine features from contracting and expanding paths %in a more complex way 
to learn more discriminative information. The dense convolutions help the network to learn more diverse features. Moreover, BN, utilized in the network, has a significant effect on the convergence speed of the network. In addition, the SE modules are exploited in the decoding path to extract more useful information by considering the interdependecies between channels of features. It is worth mentioning SE blocks are utilized in our network in a different ways than other approaches \cite{rundo2019use,zhu2018anatomynet}. Zhu et al. \cite{zhu2018anatomynet} employed SE residual block in the encoding path, and Rundo et al. utilized these blocks before the concatenation of skip connections while these blocks are inserted in the decoding path of our network.

\section{Proposed Method}
\begin{figure*}
\centering
%\vspace{-5mm}
\includegraphics[width=0.9\textwidth]{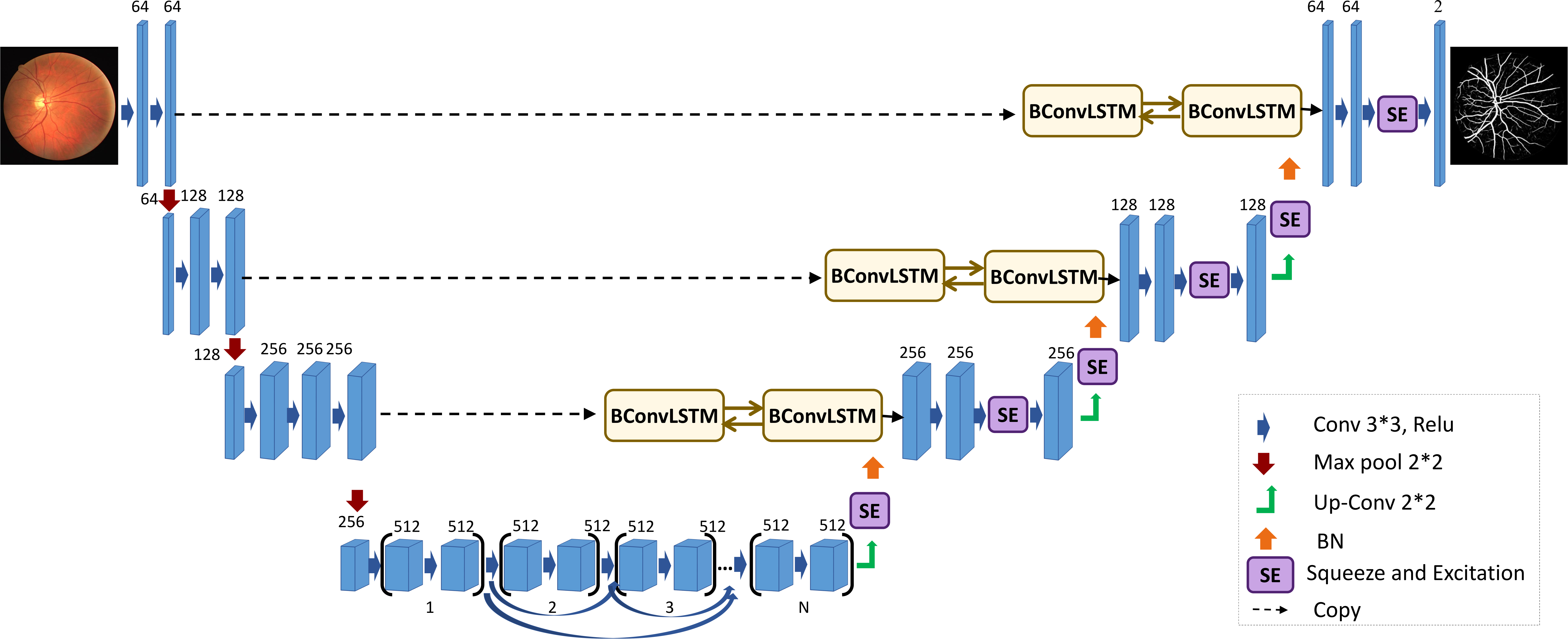}
%\vspace{-5mm}
\caption{MCGU-Net with bi-directional ConvLSTM in the skip connections, SE modlues in the decoding path, and densely connected convolution.% in the last layer of contracting path.
} \label{fig:Model}
\vspace*{-\baselineskip}
\end{figure*}

Inspired by U-Net \cite{ronneberger2015}, BConvLSTM \cite{song2018pyramid}, SENet \cite{hu2018squeeze} and dense convolutions \cite{huang2017densely}, we propose the MCGU-Net (Figure \ref{fig:Model}). %The network utilizes the strengths of all BConvLSTM states, SE modules, and densely connected convolutions. %We highlight different parts of the proposed network with more details in the follwoing sub sections.
We detail all parts of the network in the next sub sections.
%It has been proved that augmenting a convolutional layer with a self attention mechanism \cite{bello2019attention} improves the classification rate. In this paper, we add a self attention-based convolution after the last convolutional layer in the encoding path. Moreover, in skip connection, instead of a simple concatenation of the corresponding cropped feature map from the contracting path and the result of up-convolutional layer, we employed ConvLSTM to process this data with more complex functions. Moreover, the outputs of the ConvLSTM are first normalized with a batch normalization layer (BN) and then fed to the next convolutional layer. We highlight different parts of the proposed network with more details into two sub-sections of encoding and decoding paths.

%\vspace{-7mm}

\subsection{Encoding Path}
The U-Net consists of a contracting path to extract hierarchically semantic features from the input and capture context information. To improve the performance of the U-Net we utilize the idea of transfer learning by exploiting a pre-trained CNN of VGG family as the encoder \cite{van2014transfer}. To train a complex model with a huge amount of parameters, a large dataset is necessary. However, gathering a vast number of labeled data is very tough. On the other hand, deep learning models are mostly focused on a specific task.  
To overcome the isolated learning paradigm, the idea of transfer learning has been proposed, which leverage knowledge from pre-trained models and use it to solve new problem, which may have less data.  %Therefore, in the proposed MCGU-Net, we use a VGG backbone as the encoder consists of four steps. 
Inspiring by this idea, we design the encoding path like the first four layers of VGG-16 to make it possible to use pre-trained models.
The first two layers includes  two convolutional $3\times3$ filters followed by a $2\times2$ max pooling and ReLU. The number of convolutional filters in the third layer is three with the same filter size followed by the same pooling and ReLU. The number of feature maps are doubled at each step. 

%The contracting path of MCGU-Net includes four steps. Each step consists of two convolutional $3\times3$ filters followed by a $2\times2$ max pooling function and ReLU. The number of feature maps are doubled at each step. %In fact, the convolutional layers in the contracting path process the local information layer by layers. 

%The contracting path extracts progressively image representations and increases the dimension of these representations layer by layer. Ultimately, the final layer in the encoding path produces a high dimensional image representation with high semantic information.

The original U-Net contains a sequence of convolutional layers in the last step of encoding path.
Having a sequence of convolutional layers in a network yields the method learn different kinds of features. Nevertheless, the network might learn redundant features in the successive convolutions. To mitigate this problem, densely connected convolutions are proposed \cite{huang2017densely}. This helps the network to improve its performance by the idea of ``collective knowledge" in which the feature maps are reused through the network. It means feature maps learned from all previous convolutional layers are concatenated with the feature map learned from the current layer and then are forwarded to use as the input to the next convolution.

The idea of densely connected convolutions has some advantages over the regular one \cite{huang2017densely}. First, it helps the network to learn a diverse set of feature maps instead of redundant ones. Moreover, this idea improves the network's representational power by allowing information flow through the network. %and reusing features. 
Furthermore, dense connected convolutions can benefit from all the produced features before it (i.e., collecting knowledge), which prompt the network to avoid the risk of exploding or vanishing gradients. In addition, the gradients are sent to their respective places in the network more quickly in the backward path. 
We employ this idea in the proposed network. To do that, we introduce one block as two consecutive convolutions. There are a sequence of $N$ blocks in the last convolutional layer of the encoding path. %, shown in Figure \ref{fig:dense}. 
These blocks are densely connected. % to each others. 
We consider $\mathcal{X}_e^i $ as the output of the $i^{th}$ convolutional block. The input of the $i^{th}$ ($i\in \{1,...,N\}$) convolutional block receives the concatenation of the feature maps of all preceding convolutional blocks as its input, i.e., $ \big[\mathcal{X}_e^{1},\mathcal{X}_e^{2}, ...,\mathcal{X}_e^{i-1} \big] \in \mathbb{R}^{(i-1)F_l\times W_l\times H_l}$ , and the output of the $i^{th}$ block is $\mathcal{X}_e^{i} \in \mathbb{R}^{F_l\times W_l\times H_l}$. In the remaining part of the paper we use simply $\mathcal{X}_e$ instead of $\mathcal{X}_e^{N}$.

% \begin{figure*}
% \centering
% %\vspace{-5mm}
% \includegraphics[width=0.8\textwidth]{Dense2.pdf}
% %\vspace{-5mm}
% \caption{Dense layer of the MCGU-Net.} \label{fig:dense}
% \vspace*{-\baselineskip}
% \end{figure*}

%After the last convolutional layer, a self-attention based convolutional layer is included. The purpose of self-attention is to dynamically produce a weighted average of values computed from hidden units via a similarity function between them. Bello et al. \cite{bello2019attention} proposed to augment a convolutional layer with a self-attention, which maintains translation equivariance while being infused with relative position information. To do that, the convolutional feature maps are concatenated by the self-attention feature maps. %A number of attention maps (multihead attentions) are computed from a set of queries and keys for each spatial location of the input data. The output of these attention maps are a number of weighted averages of the values which are then concatenated and reshaped to match the original spatial size. The convolution is performed in parallel to multihead attention, and its feature maps are concatenated with the output of multihead attention. More information about the convolutional attention can be found in \cite{bello2019attention}.

\subsection{Decoding Path}
\begin{figure}
\centering
%\vspace{-5mm}
\includegraphics[width=0.45 \textwidth]{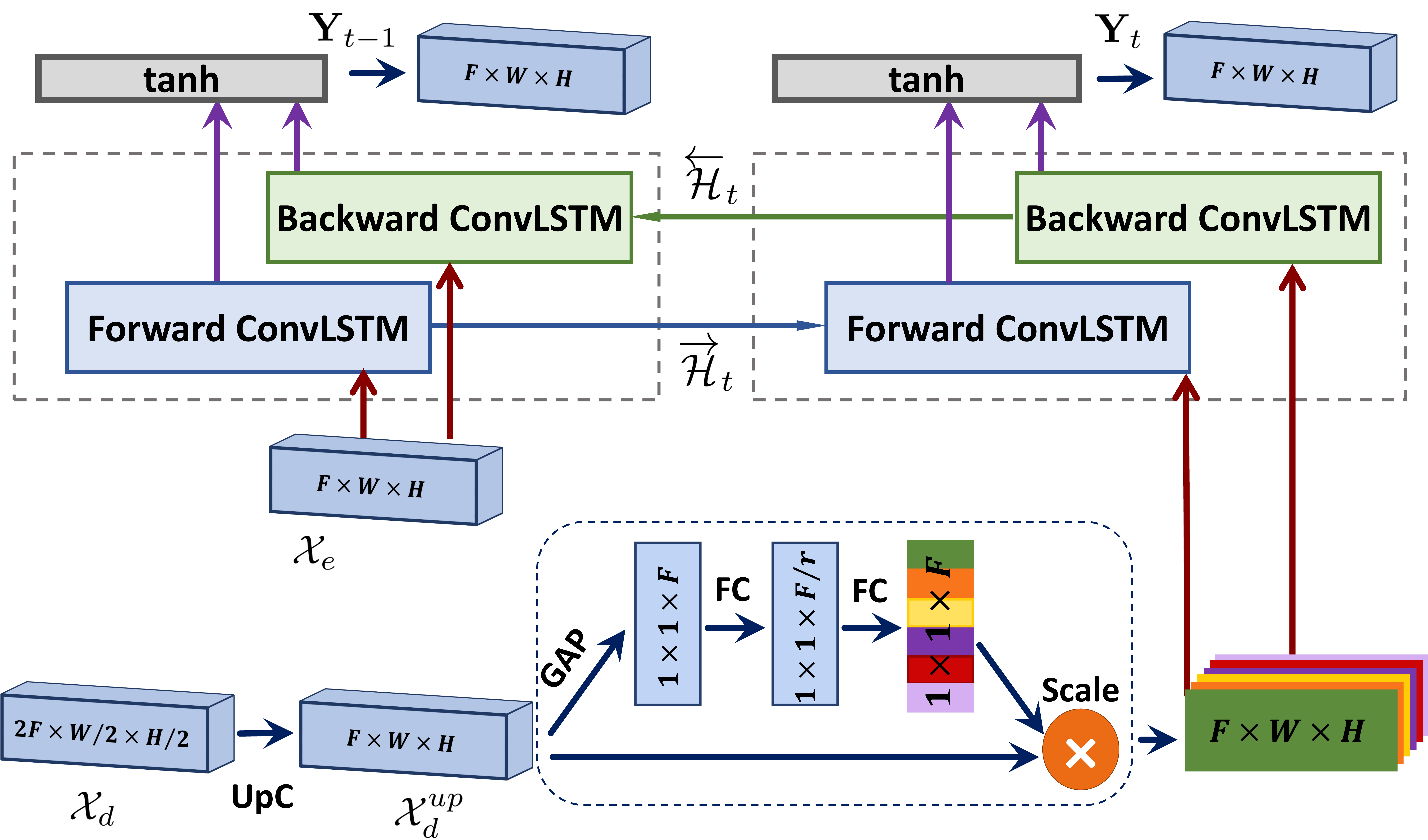}
%\vspace{-4mm}
\caption{BConvLSTM with SE block in the decoding path of MCGU-Net.} \label{fig:BConvLSTM}
\vspace*{-\baselineskip}
\end{figure}

Each step in the decoding path starts with an up-sampling function over the output of the previous layer. To improve the representation power of the network, decoding path of the original U-Net is augmented with two important modules of SE block and BConvLSTM. 
SE yield the network to use global information to selectively empathize informative features and suppress less useful ones. This block receives the output of the up-sampling function, which is a collection of feature maps, and encourages the feature maps to be more informative using a weight for each channel based on the interdependencies between all channels. 
The output of the SE module is then passed to an up-sampling function. In the standard U-Net, the corresponding feature maps in the contracting path are concatenated with the output of the up-sampling function. In the proposed network, we employ BConvLSTM to combine these two kinds of feature maps. The output of the BConvLSTM is then fed to a set functions including two convolutional functions, one SE module, and another convolutional filter. A diagram illustrating the structure of the combination of these modules in our network is shown in Figure \ref{fig:BConvLSTM}.

Assume that the set of extracted feature maps from the previous layer in decoding path is $\mathcal{X}_d\in \mathbb{R}^{F_{l+1}\times W_{l+1}\times H_{l+1}}$ where $F_l$ is the number of feature maps at layer $l$, and $W_l\times H_l$ is the size of each feature map at layer $l$. We have $F_{l+1} = 2 *F_{l}$, $W_{l+1} = \frac{1}{2} *W_l$, and $ H_{l+1} = \frac{1}{2} * H_l$. For simplicity we consider $\mathcal{X}_d\in \mathbb{R}^{2F\times \frac{W}{2}\times \frac{H}{2}}$. As it can be seen in Figure \ref{fig:BConvLSTM}, this set of feature maps is first passed through an up-convolutional layer in which an up-sampling function followed by a $2\times2$ convolution are applied, doubling the size of each feature map and halving the number of channels, i.e., producing  $\mathcal{X}_d^{up}\in \mathbb{R}^{F\times W\times H}$. In other words, the expanding path increases the size of the feature maps layer by layer to reach the original size of the input image after the final layer.

%Based on Figure \ref{fig:BConvLSTM}, $\mathcal{X}_d$  is first passed to an up-convolutional layer in which an up-sampling function followed by a $2\times2$ convolution are applied, doubling the size of each feature map and halving the number of feature channels, i.e., producing $\mathcal{X}_{d}^{up}\in \mathbb{R}^{F_l\times W_l\times H_l}$. In other words, the expanding path increases the size of the feature maps layer by layer to reach the original size of the input image after the final layer.

%Each step in the decoding path starts with performing an up-sampling function over the output of the previous layer. In the standard U-Net, the corresponding feature maps in the contracting path are cropped and copied to the decoding path. These feature maps are then concatenated with the output of the up-sampling function.

\subsubsection{Squeeze and Excitation Module}
Capturing spatial correlations between features has improved the performance of deep networks, like Inception architectures \cite{szegedy2015going} and spatial attention \cite{jaderberg2015spatial}. 
%However, these networks weight each of the channels similarly when creating the output feature maps. 
However, the network produces the same attention to the channels when creating the output feature maps.
The SE mechanism \cite{hu2018squeeze} is proposed to capture explicit relationship between channels of the convolutional layers by a context gating mechanism, which results in improving the representation power of the network. These modules encode feature maps by assigning a weight for each channel (i.e. channel attention) in the feature map.

The SE block includes two parts squeeze and excitation. The first operation is squeeze. The input feature maps to SE block are aggregated to generate channel descriptor by employing global average pooling (GAP) of the whole context of channels. We have  $\mathcal{X}_d^{up}=[x_1^{up},x_2^{up},...,x_F^{up}]$, where $x_f^{up} \in \mathbb{R}^{W\times H}$, as the input data to the SE block. The spatial squeeze (GAP) is performed as
\begin{equation}
    z_f = \textbf{F}_{sq}(x_f^{up}) =  \frac{1}{H\times W}\sum_{i}^{H} \sum_{j}^{W} x_f^{up}(i,j)
\end{equation}

\noindent where $\textbf{F}_{sq}$ are the spatial squeeze function, $x_f^{up}(i,j)$ is a spatial location of the $f^{th}$ channel, and $H\times W$ is the size of this channel. In other words, each two-dimensional feature map is compressed by a global average pooling to produce $z_f$. 
To capture the channel-wise dependencies, the global information embedded in the first step is then fed to the second step, i.e., Excitation. This function must be able to learn non-mutually-exclusive relationship and nonlinear interaction between channels \cite{hu2018squeeze}. As it is illustrated in Figure \ref{fig:BConvLSTM}, the excitation step consists of two fully connected (FC) layers. The pooled vector is first encoded to shape $1\times 1 \times \frac{F}{r}$, and then encoded again to shape $1\times 1 \times F$ to generate excitation vector as 
\begin{equation}
   \textbf{s} = \textbf{F}_{ex}(\textbf{z};\textbf{W}) = \sigma\left( \textbf{W}_2 \delta(\textbf{W}_1 \textbf{z}) \right)
\end{equation}

\noindent where $\textbf{W}_1 \in \mathbb{R}^{\frac{F}{r}\times F}$ is the parameters of the first fc layer, $\mathbb{R}^{F \times \frac{F}{r}}$, $\delta$ is ReLU, and the $\sigma$ refers to the sigmoid activation. Moreover, $r$ is the reduction ratio. In \cite{hu2018squeeze}, to limit model complexity and aid generalization, a dimensionality-reduction layer with reduction ratio $r$ is used in the first fc layer. In the second fc layer a dimensionality-increasing layer is utilized to set the dimension to the channel one of the transformation output. 
The output of the SE block is generated as $\tilde{x}_f^{up} = \textbf{F}_{scale}(x_f^{up}, z_c) = s_c\  x_f^{up}$. In which $\tilde{\mathcal{X}}_d^{up} = [\tilde{x}_1^{up}, \tilde{x}_2^{up}, ...,\tilde{x}_F^{up}]$, and $\textbf{F}_{scale}$ is a channel-wise multiplication between the channel attention, the scale factor $s_c$, and the input feature map.
%\begin{equation}
%   \tilde{x}_f^{up} = \textbf{F}_{scale}(x_f^{up}, z_c) = s_c\  x_f^{up}.
%\end{equation}

%\noindent where $\tilde{\mathcal{X}}_d^{up} = [\tilde{x}_1, \tilde{x}_2, ...,\tilde{x}_F]$, and $\textbf{F}_{scale}$ is a channel-wise multiplication between the scale factor $s_c$ and the input feature map.
%

%In MCGU-Net, we employ BConvLSTM to process these two kinds of feature maps in a more complex way. 

%Let $\mathcal{X}_e \in \mathbb{R}^{F_l\times W_l\times H_l}$ be the set of feature maps copied from the encoding part, and $\mathcal{X}_d\in \mathbb{R}^{F_{l+1}\times W_{l+1}\times H_{l+1}}$ be the the set of feature maps from the previous convolutional layer, where $F_l$ is the number of feature maps at layer $l$, and $W_l\times H_l$ is the size of each feature map at layer $l$. It is worth mentioning that $F_{l+1} = 2 *F_{l}$, $W_{l+1} = \frac{1}{2} *W_l$, and $ H_{l+1} = \frac{1}{2} * H_l$. 

%\vspace*{-\baselineskip}
\subsubsection{Batch Normalization}
After up-sampling, $\tilde{\mathcal{X}}_d^{up}$ goes through a BN function and produces $\widehat{\mathcal{X}}_{d}^{up}$. A problem in the intermediate layers in training step is that the distribution of the activations varies. This problem makes the training process very slow since each layer in every training step has to learn to adapt themselves to a new distribution. BN \cite{ioffe2015batch} is utilized to increase the stability of a neural network, which standardizes the inputs to a layer in the network by subtracting the batch mean and dividing by the batch standard deviation. BN affectedly accelerates the speed of training process of a neural network. Moreover, in some cases the performance of the model is improved thanks to the modest regularization effect. More details can be found in \cite{ioffe2015batch}.

\subsubsection{Bi-Directional ConvLSTM}
The output of the BN step ($\widehat{\mathcal{X}}_{d}^{up} \in \mathbb{R}^{F_l\times W_l\times H_l}$) is now fed to a BConvLSTM layer. 
The main disadvantage of the standard LSTM is that these networks does not take into account the spatial correlation since these models use full connections in input-to-state and state-to-state transitions. To solve this problem, ConvLSTM \cite{xingjian2015} was proposed which exploited convolution operations into input-to-state and state-to-state transitions. It consists of an input gate $i_t$, an output gate $o_t$, a forget gate $f_t$, and a memory cell $\mathcal{C}_t$. Input, output and forget gates act as controlling gates to access, update, and clear memory cell. ConvLSTM can be formulated as follows (for convenience we remove the subscript and superscript from the parameters):

%\vspace*{-\baselineskip}
\begin{equation}
\label{equ:ConvLSTM1}
\centering
\begin{aligned}
    &i_t = \sigma \left( \textbf{W}_{xi} * \mathcal{X}_t + \textbf{W}_{hi} * \mathcal{H}_{t-1} + \textbf{W}_{ci} * \mathcal{C}_{t-1} + b_i \right) \\
    &f_t = \sigma \left( \textbf{W}_{xf}* \mathcal{X}_t + \textbf{W}_{hf}* \mathcal{H}_{t-1} +\textbf{W}_{cf}* \mathcal{C}_{t-1} + b_f \right) \\
    &\mathcal{C}_t = f_t \circ \mathcal{C}_{t-1} + i_t \tanh\left( \textbf{W}_{xc}* \mathcal{X}_t + \textbf{W}_{hc}* \mathcal{H}_{t-1} + b_c \right) \\
    & o_t = \sigma \left( \textbf{W}_{xo}* \mathcal{X}_t + \textbf{W}_{ho}* \mathcal{H}_{t-1} +\textbf{W}_{co}\circ \mathcal{C}_{t} + b_c \right) \\
    &\mathcal{H}_t = o_t \circ \tanh(\mathcal{C}_t),\\
\end{aligned}
\end{equation}
%\vspace*{-\baselineskip}

\noindent where $*$ and $\circ$ denote the convolution and Hadamard functions, respectively. $\mathcal{X}_t$ is the input tensor (in our case $\mathcal{X}_e$ and $\widehat{\mathcal{X}}_{d}^{up}$), $\mathcal{H}_t$ is the hidden sate tensor, $\mathcal{C}_t$ is the memory cell tensor, and, $\textbf{W}_{x*}$ and $\textbf{W}_{h*}$ are 2D Convolution kernels corresponding to the input and hidden state, respectively, and $b_i$, $b_f$, $b_o$, and $b_c$ are the bias terms.

In this network, we employ BConvLSTM \cite{song2018pyramid} to encode $\mathcal{X}_e $ and $\widehat{\mathcal{X}}_{d}^{up}$. BConvLSTM uses two ConvLSTMs to process the input data into two directions of forward and backward paths, and then makes a decision for the current input by dealing with the data dependencies in both directions. In a standard ConvLSTM, only the dependencies of the forward direction are processed. However, all the information in a sequence should be fully considered, therefore, it might be effective to take into account backward dependencies. 
It has been proved that analyzing both forward and backward temporal perspectives enhanced the predictive performance \cite{cui2018deep}.
Each of the forward and backward ConvLSTM can be considered as a standard one. Therefore, we have two sets of parameters for backward and forward states. 
The output of the BConvLSTM is calculated as $\textbf{Y}_t = \tanh{\left( \textbf{W}_y^{\overrightarrow{\mathcal{H}}} * \overrightarrow{\mathcal{H}}_t +   \textbf{W}_y^{\overleftarrow{\mathcal{H}}} \overleftarrow{\mathcal{H}}_t + b \right)}$. 
In which $\overrightarrow{\mathcal{H}}_t$ and $\overleftarrow{\mathcal{H}}_t$ denote the hidden sate tensors for forward and backward states, respectively, $b$ is the bias term, and $\textbf{Y}_t \in \mathbb{R}^{F_l\times W_l\times H_l}$ indicates the final output considering bidirectional spatio-temporal information. Moreover, $\tanh$ is the hyperbolic tangent which is utilized here to combine the output of both forward and backward states through a non-linear way. We utilize the energy function like the original U-Net to train the network.

\section{Experimental result}
We evaluate MCGU-Net on six datasets of: DRIVE, ISIC 2017, ISIC 2018, a lung segmentation benchmark, $PH^2$, and a cell nuclei segmentation dataset. %DRIVE is for blood vessel segmentation from retina images. Both ISICs and $PH^2$ datasets are for the skin cancer lesion segmentation. The lung segmentation one consists of diagnostic and lung cancer screening thoracic CT scans with marked-up annotated lesions. The last dataset is a part of Data Science Bowl 2018 for cell nuclei segmentation. 
The empirical results show that the proposed method outperforms state-of-the-art alternatives for all six datasets. Keras with TenserFlow backend is utilized for implementation. %It is worth mentioning that 
%The network is trained from scratch for all datasets.
We consider several performance metrics to perform the experimental comparative, including accuracy (AC), sensitivity (SE), specificity (SP), F1-Score, Jaccard similarity (JS), and area under the curve (AUC). We stop the training of the network when the validation loss remains the same in 10 consecutive epochs.% which is 50, 100, and 25 for DRIVE, ISIC, and Lung datasets, respectively.
%By using variables TP (true positive), FP (false positive), TN (true negative), and FN (false negative), GT (ground truth) and SR (segmentation result) evaluation metrics are calculated as 

% \begin{equation}
% \label{equ:measure}
% \centering
% \begin{aligned}
% &AC = \frac{TP+TN}{TP+FP+TN+FN} \\
% &SE = \frac{TP}{TP+FN} \\
% &SP = \frac{TN}{TN+FP} \\
% &PC = \frac{TP}{TP+FP}\\
% &F1 =2* \frac{PC*SE}{PC+SE}\\
% &JS = \frac{|GR\cap SR|}{|GR\cup SR|}\\
% \end{aligned}
% \end{equation}

% \begin{figure*}[ht]
% 	\centering
% 	\begin{tabular}{ccc}
% 		% Requires \usepackage{graphicx}
% 		\includegraphics[width=0.3\textwidth]{DRIVE_Dataset.pdf}&
% 		\includegraphics[width=0.3\textwidth]{ISIC_Dataset.pdf}&
% 		\includegraphics[width=0.3\textwidth]{Lung_Dataset.pdf}\\
% 		(a) DRIVE, & (b) ISIC, & (c) Lung Segmentation,\\
% 	\end{tabular}
% 	\caption{Some samples of three datasets.}
% 	\label{fig:Datasets}
% \end{figure*}

\subsection{DRIVE Dataset}

DRIVE \cite{staal2004} is a dataset for blood vessel segmentation from retina images. %, shown in Figure \ref{fig:Datasets} (a).
It includes 40 color retina images, from which 20 samples are used for training and the remaining 20 samples for testing. The original size of images is $565\times584$ pixels. It is clear that a dataset with this number of samples is not sufficient for training a deep network. Therefore, we use the same strategy as \cite{alom2018} for training our network. The input images are first randomly divided into a number of patches ($64\times64$). In total, around $190,000$ patches are produced from $20$ training images, from which $171,000$ patches are used for training, and the remaining $19,000$ patches are used for validation. % The size of batches utilized as the input data to the network is $64\times64$. 

% \begin{figure}
% \centering
% \includegraphics[width=0.5\textwidth]{DRIVE_Dataset.pdf}
% \caption{Some samples of DRIVE dataset.} \label{fig3DRIVE}
% \end{figure}

\begin{figure}
\centering
\includegraphics[width=0.38\textwidth]{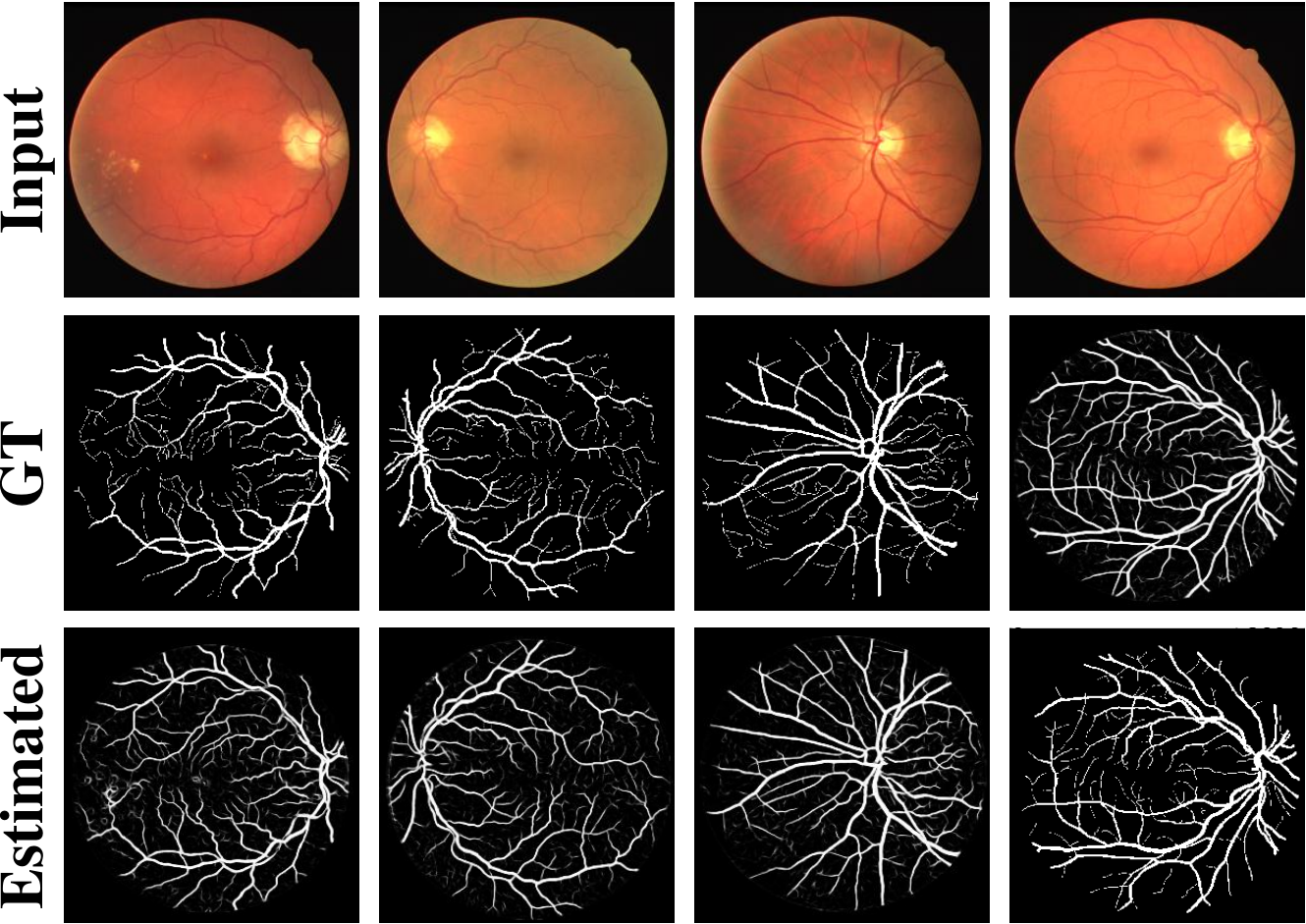}
\caption{Segmentation result of MCGU-Net on DRIVE.} 
\vspace*{-0.5\baselineskip}
\label{fig:DRIVE_R}
\end{figure}

Some precise and promising segmentation results of the %experimental output of the 
proposed network are shown in Figure \ref{fig:DRIVE_R}. %The first column is the original color image, the second one is the ground truth mask and the third column is the output of the proposed MCGU-Net. 
%The training and validation accuracy of the proposed network is shown in Figure \ref{fig:converge} (a) for DRIVE dataset. It can be seen that he network is converged very fast for this dataset.
Table \ref{tab:drive} lists the quantitative results obtained by other methods and the proposed network on DRIVE dataset. %Term "d" in this table means the number of dense block we utilized in the network. 
We evaluate the network with $d=1$ and $d=3$ as the number of dense blocks. % in the network. 
With $d=1$ we have one convolutional block without any dense connection in that layer. %, i.e., like the last encoding layer of the standard U-Net. 
With $d=3$ we have three convolutional blocks and two dense connections in that layer. 
It is shown that the MCGU-Net (with both $d=1$ and $d=3$) outperforms w.r.t. the state-of-the-art alternatives for most of the evaluation metrics. Moreover, it can be seen that the network with $d=3$ works better than $d=3$. %the network without dense block.
It is worth mentioning that for this dataset, we achieved better result by training the network from scratch rather than using pre-trained weights.

%\begin{table*}
%\centering
%    \vspace*{-\baselineskip}
%	\caption{Performance comparison of the proposed network and the state-of-the-art methods on DRIVE dataset.}
%	\begin{tabular}{cccccc}
%		\hline
%		\textbf{Methods} & \textbf{F1-Score}&	\textbf{Sensitivity}&	\textbf{Specificity}&	\textbf{Accuracy}&	\textbf{AUC}\\
%		\hline
%		%Hybrid Features \cite{cheng2014} &  -	&0.7252&	0.9798&	0.9474&	0.9648\\
%		%Three Stage Filtering \cite{roychowdhury2014} & -&	0.7250&	\textbf{0.9830}&	0.9520&	0.9620\\
%		%COSFIRE filters \cite{azzopardi2015} & -&	0.7655&	0.97048&	0.9442&	0.9614\\
%		%Cross-Modality \cite{li2015cross} & -&	0.7569&	0.9816&	0.9527&	0.9738\\
%		U-net \cite{ronneberger2015} & 0.8142&	0.7537&	0.9820&	0.9531&	0.9755\\
%		Deep Model \cite{liskowski2016} & -&	0.7763&	0.9768&	0.9495&	0.9720\\
%		Attention U-net \cite{oktay2018} & 0.8155&	0.7751&	0.9816&	0.9556&	0.9782\\
%		RU-net \cite{alom2018} & 0.8149&	0.7726&	0.\textbf{9820}&	0.9553&	0.9779\\
%		R2U-Net \cite{alom2018} &0.8171& 0.7792&	0.9813&	0.9556&	0.9782\\
%		%BCDU-Net (d=1) \cite{azad2019bi}& 0.8222& 0.8012&0.9784&	0.9559&  0.9788\\
%		%BCDU-Net (d=3) \cite{azad2019bi}& 0.8224&0.8007&0.9786&	0.9560& 0.9789\\
%		%BCDU-Net \cite{azad2019bi}& 0.8224&0.8007&0.9786&	0.9560& 0.9789\\
%		\hline
%		\textbf{MCGU-Net (d=1)} & 0.8222& \textbf{0.8012}& 0.9784 &0.9559 & 0.9788\\
%		\textbf{MCGU-Net (d=3)} & \textbf{0.8224} &0.8007& \textbf{0.9786}&	\textbf{0.9560}& \textbf{0.9789}\\
%		\hline
%	\end{tabular}
%	\label{tab:drive}
%\end{table*}
%%\vspace*{-\baselineskip}  

\begin{table}
\centering
    \vspace*{-\baselineskip}
	%\caption{Performance comparison of the proposed network and the state-of-the-art methods on DRIVE dataset.}
	\caption{Performance comparison on DRIVE dataset.}
	\begin{tabular}{cccccc}
		\hline
		\textbf{Methods} & \textbf{F1}&	\textbf{SE}&	\textbf{SP}&	\textbf{AC}&	\textbf{AUC}\\
		\hline
		%Hybrid Features \cite{cheng2014} &  -	&0.7252&	0.9798&	0.9474&	0.9648\\
		%Three Stage Filtering \cite{roychowdhury2014} & -&	0.7250&	\textbf{0.9830}&	0.9520&	0.9620\\
		%COSFIRE filters \cite{azzopardi2015} & -&	0.7655&	0.97048&	0.9442&	0.9614\\
		%Cross-Modality \cite{li2015cross} & -&	0.7569&	0.9816&	0.9527&	0.9738\\
		U-net \cite{ronneberger2015} & 0.8142&	0.7537&	0.9820&	0.9531&	0.9755\\
		Deep Model \cite{liskowski2016} & -&	0.7763&	0.9768&	0.9495&	0.9720\\
		Att U-net \cite{oktay2018} & 0.8155&	0.7751&	0.9816&	0.9556&	0.9782\\
		RU-net \cite{alom2018} & 0.8149&	0.7726&	0.\textbf{9820}&	0.9553&	0.9779\\
		R2U-Net \cite{alom2018} &0.8171& 0.7792&	0.9813&	0.9556&	0.9782\\
		%BCDU-Net (d=1) \cite{azad2019bi}& 0.8222& 0.8012&0.9784&	0.9559&  0.9788\\
		%BCDU-Net (d=3) \cite{azad2019bi}& 0.8224&0.8007&0.9786&	0.9560& 0.9789\\
		%BCDU-Net \cite{azad2019bi}& 0.8224&0.8007&0.9786&	0.9560& 0.9789\\
		\hline
		\textbf{MCGU-Net (d=1)} & 0.8222& \textbf{0.8012}& 0.9784 &0.9559 & 0.9788\\
		\textbf{MCGU-Net (d=3)} & \textbf{0.8224} &0.8007& \textbf{0.9786}&	\textbf{0.9560}& \textbf{0.9789}\\
		\hline
	\end{tabular}
	\label{tab:drive}
\end{table}
%\vspace*{-\baselineskip}   

\begin{figure}[ht]
	\centering
	\begin{tabular}{cc}
		\includegraphics[width=0.22\textwidth,height=25mm]{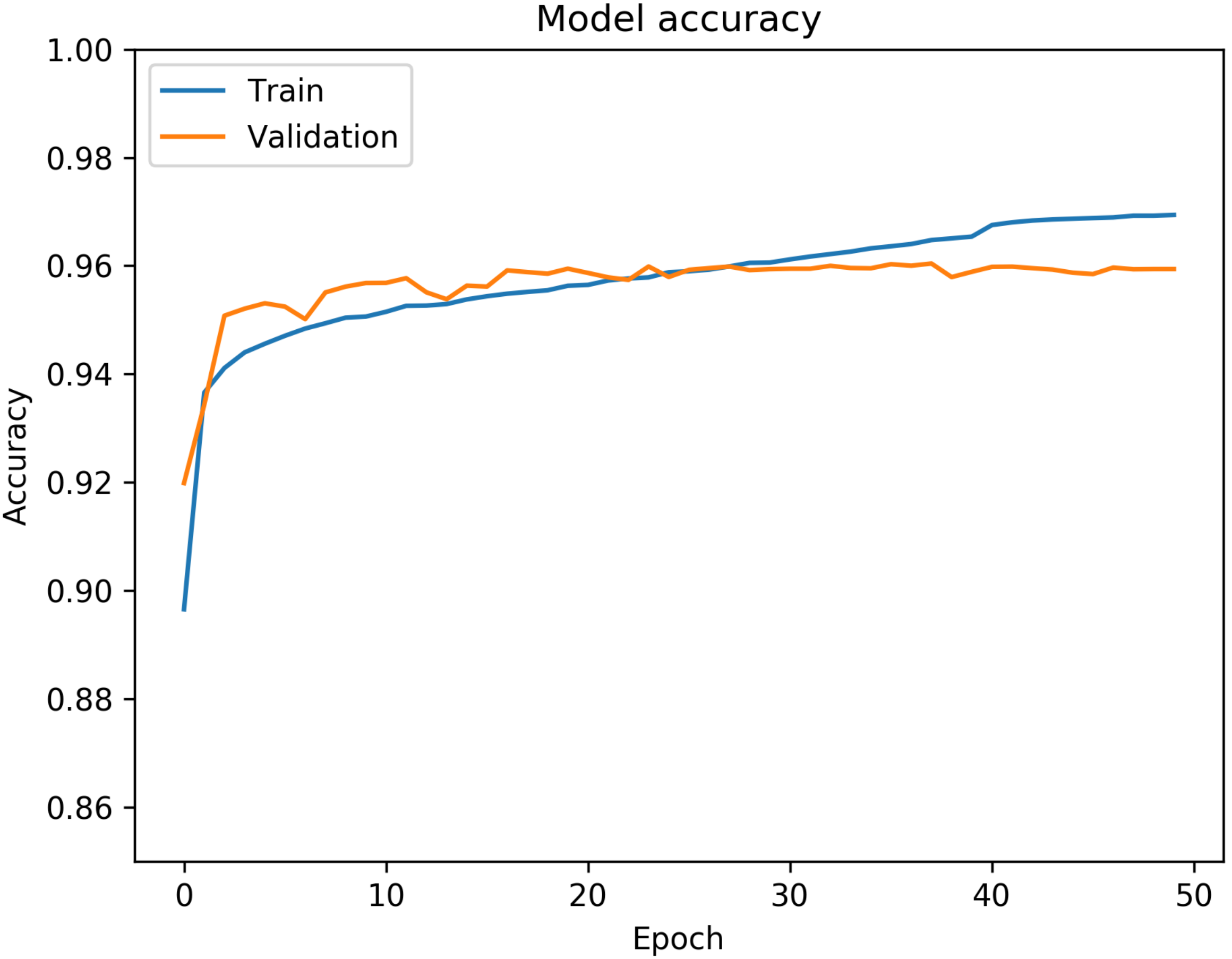}&
		\includegraphics[width=0.22\textwidth,height=25mm]{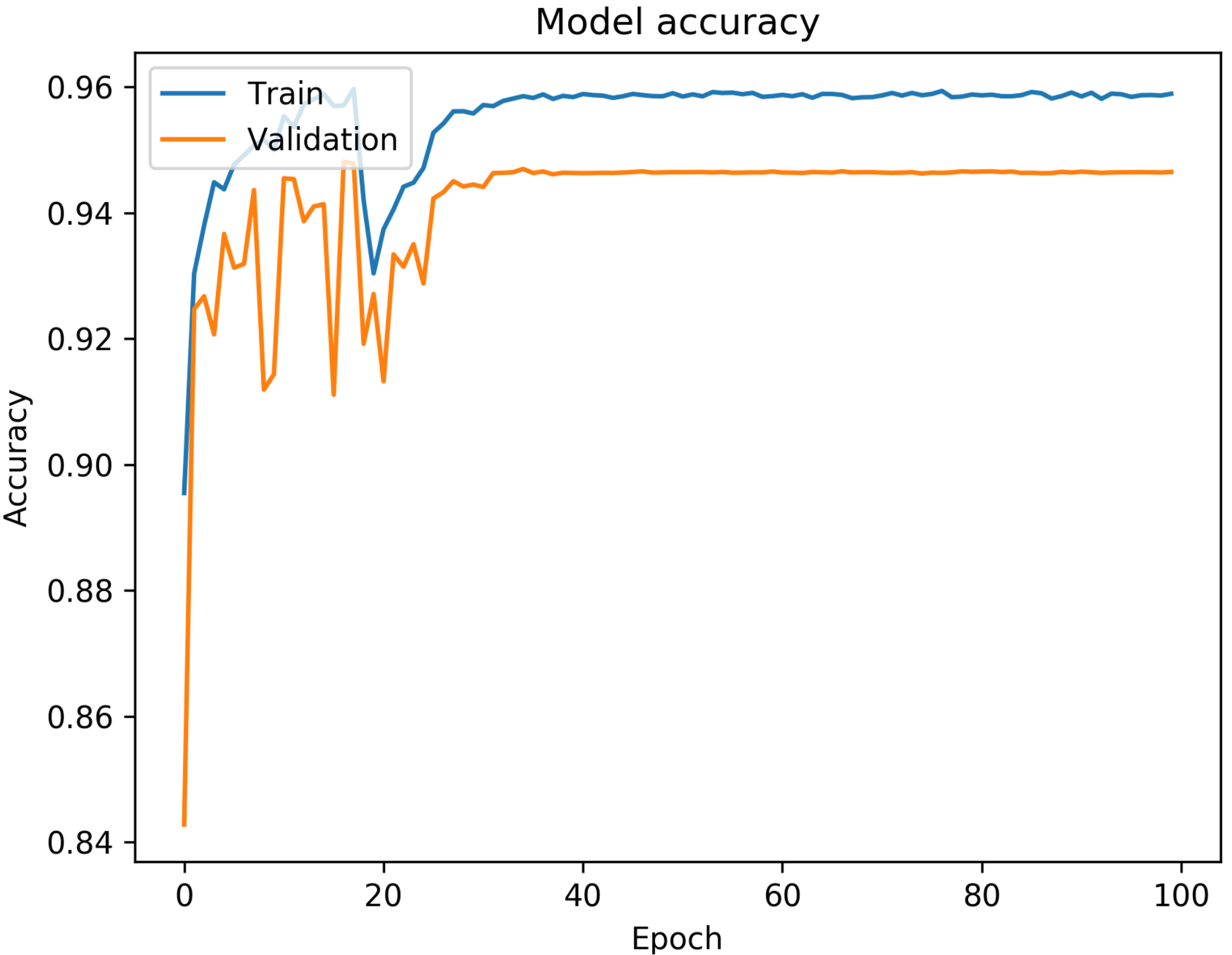} \\
		(a) DRIVE, & (b) ISIC 2017, \\
		\includegraphics[width=0.22\textwidth,height=25mm]{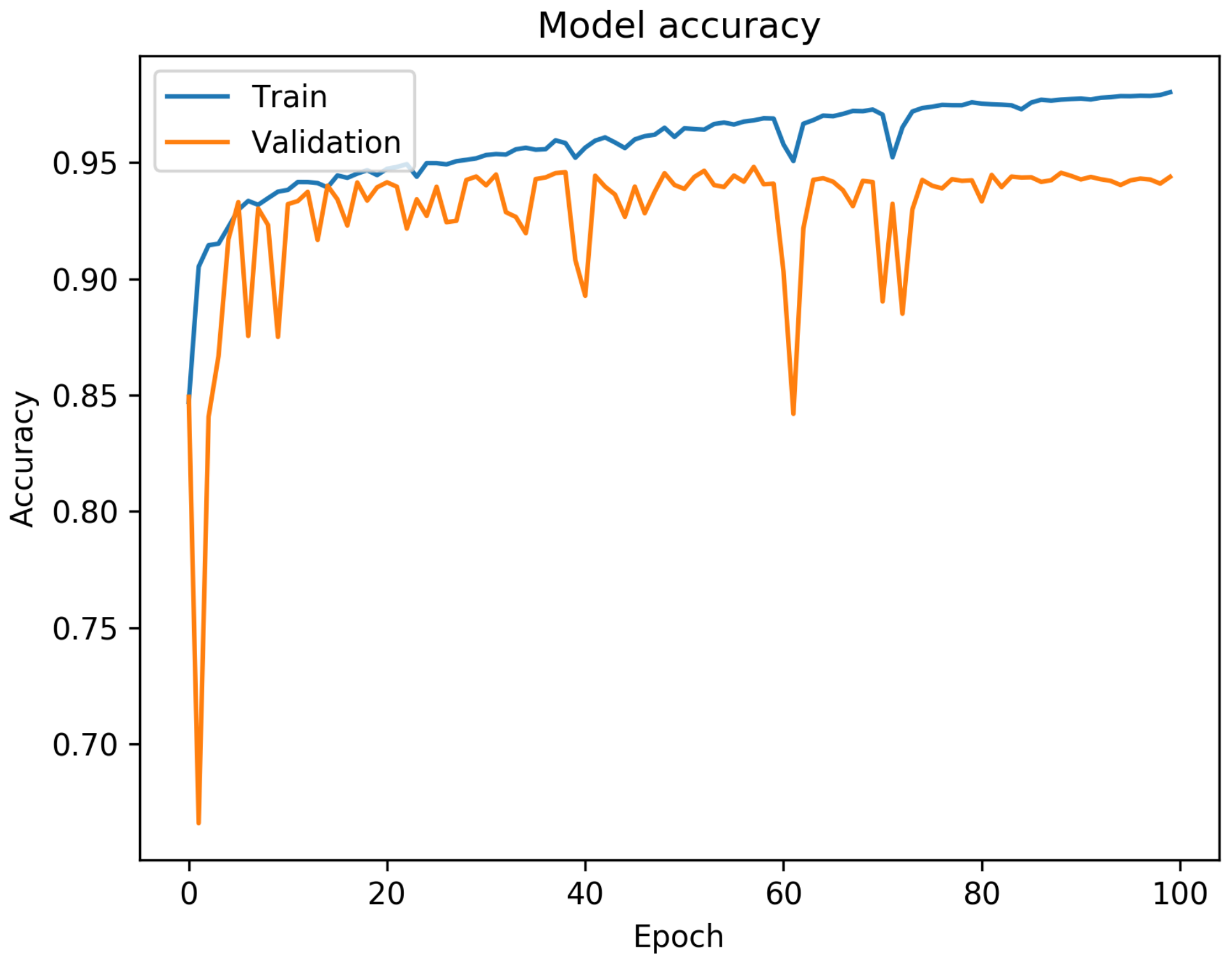} &
		\includegraphics[width=0.22\textwidth,height=25mm]{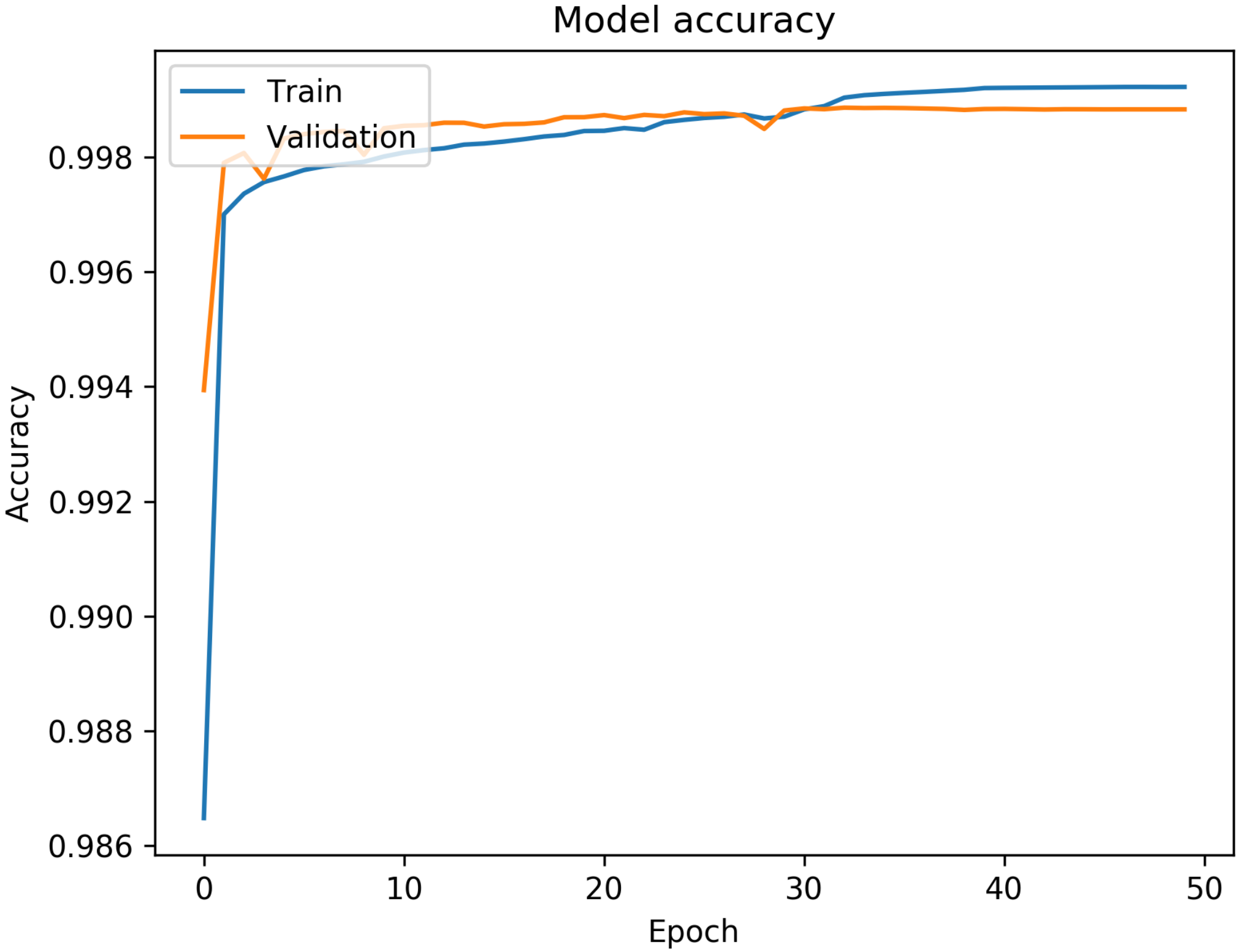} \\
		(c) ISIC 2018,&  (d) Lung Segmentation,  \\
		\includegraphics[width=0.22\textwidth,height=25mm]{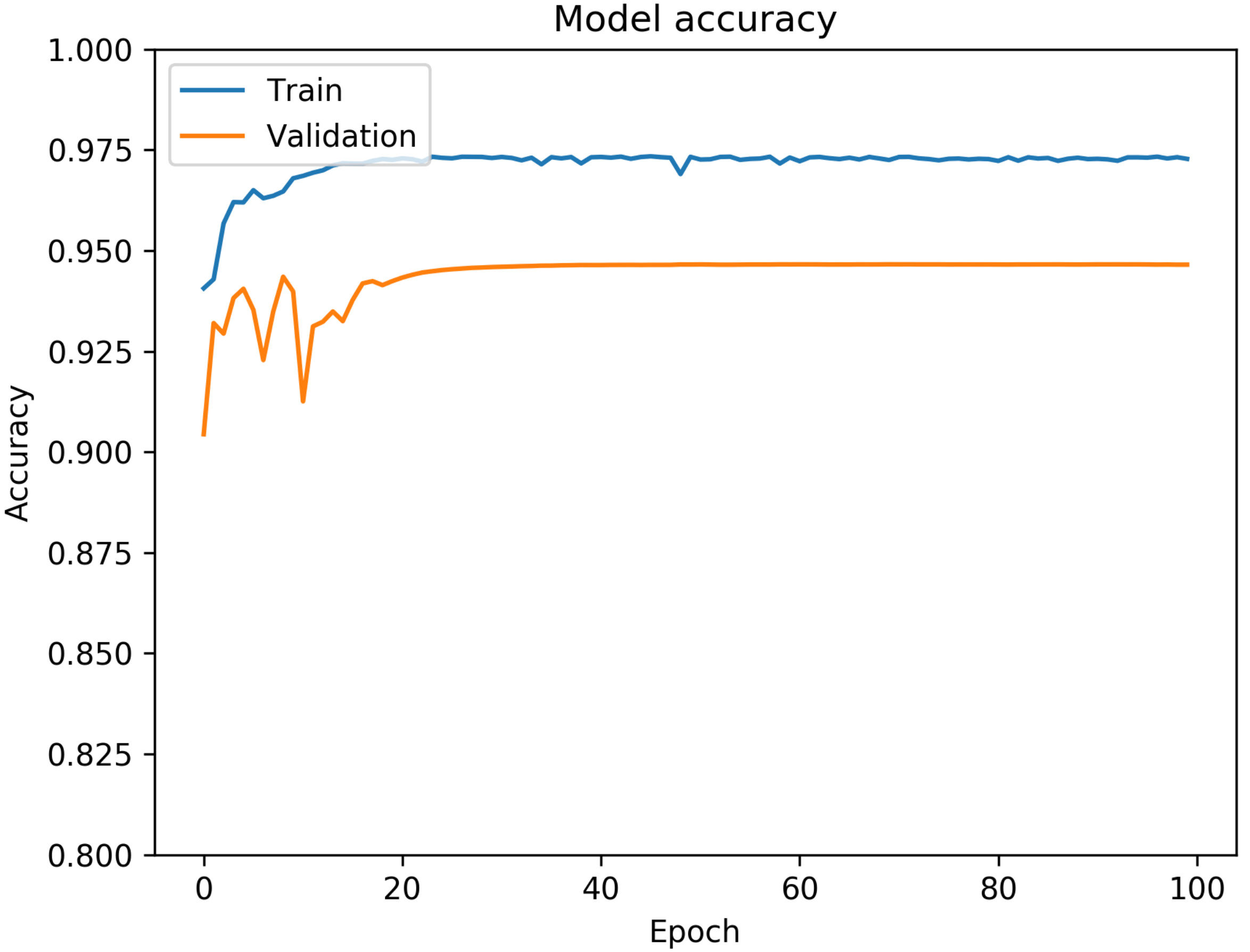} &
		\includegraphics[width=0.22\textwidth,height=25mm]{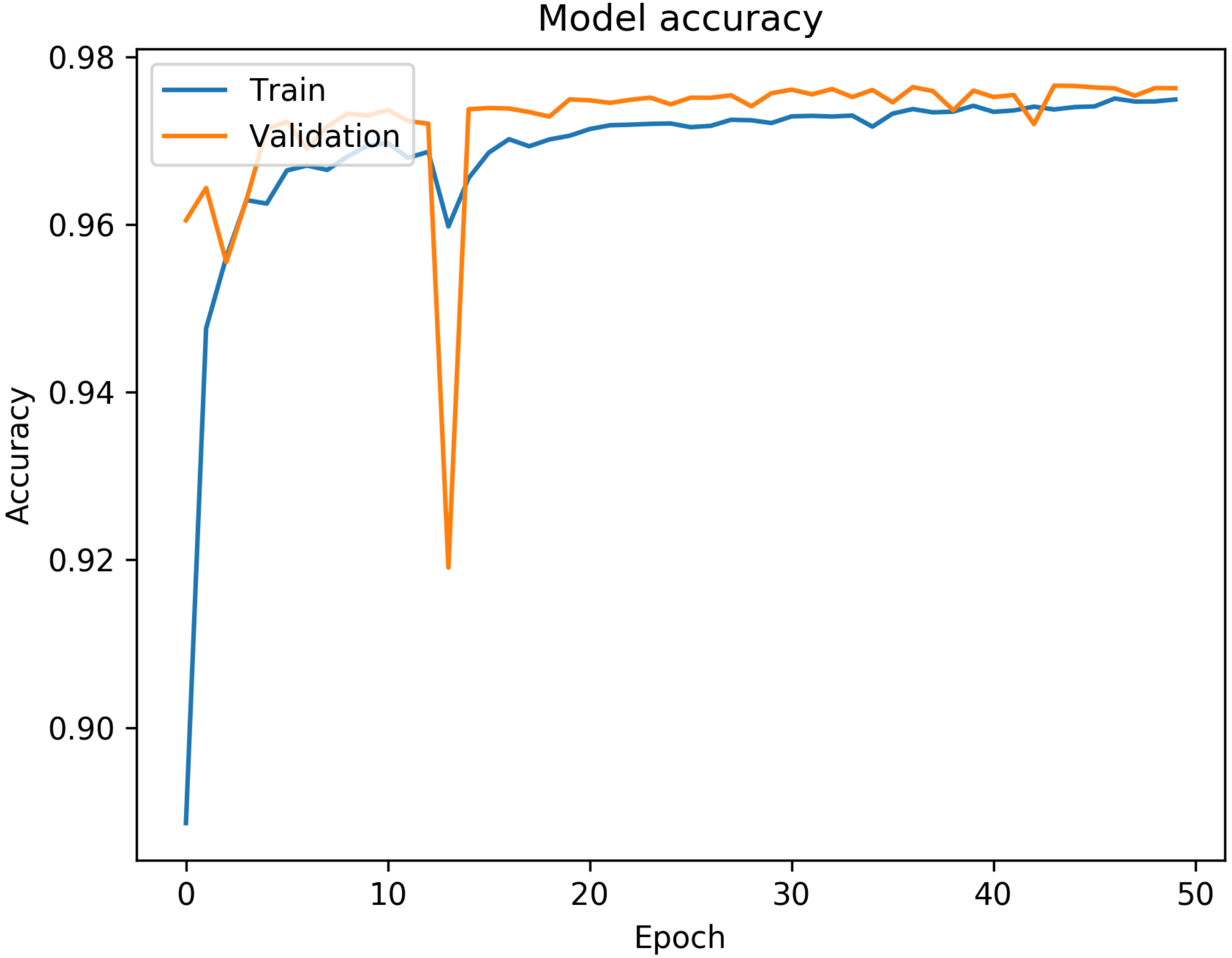}  \\
		 (e) $PH^2$,  & (f) Cell Nuclei Dataset\\
	\end{tabular}
	\caption{Training and validation accuracy of MCGU-Net for six datasets.}
	\vspace*{-\baselineskip}
	\label{fig:converge}
\end{figure}

\begin{figure}[ht]
	\centering
	\begin{tabular}{cc}
		\includegraphics[width=0.22\textwidth,height=25mm]{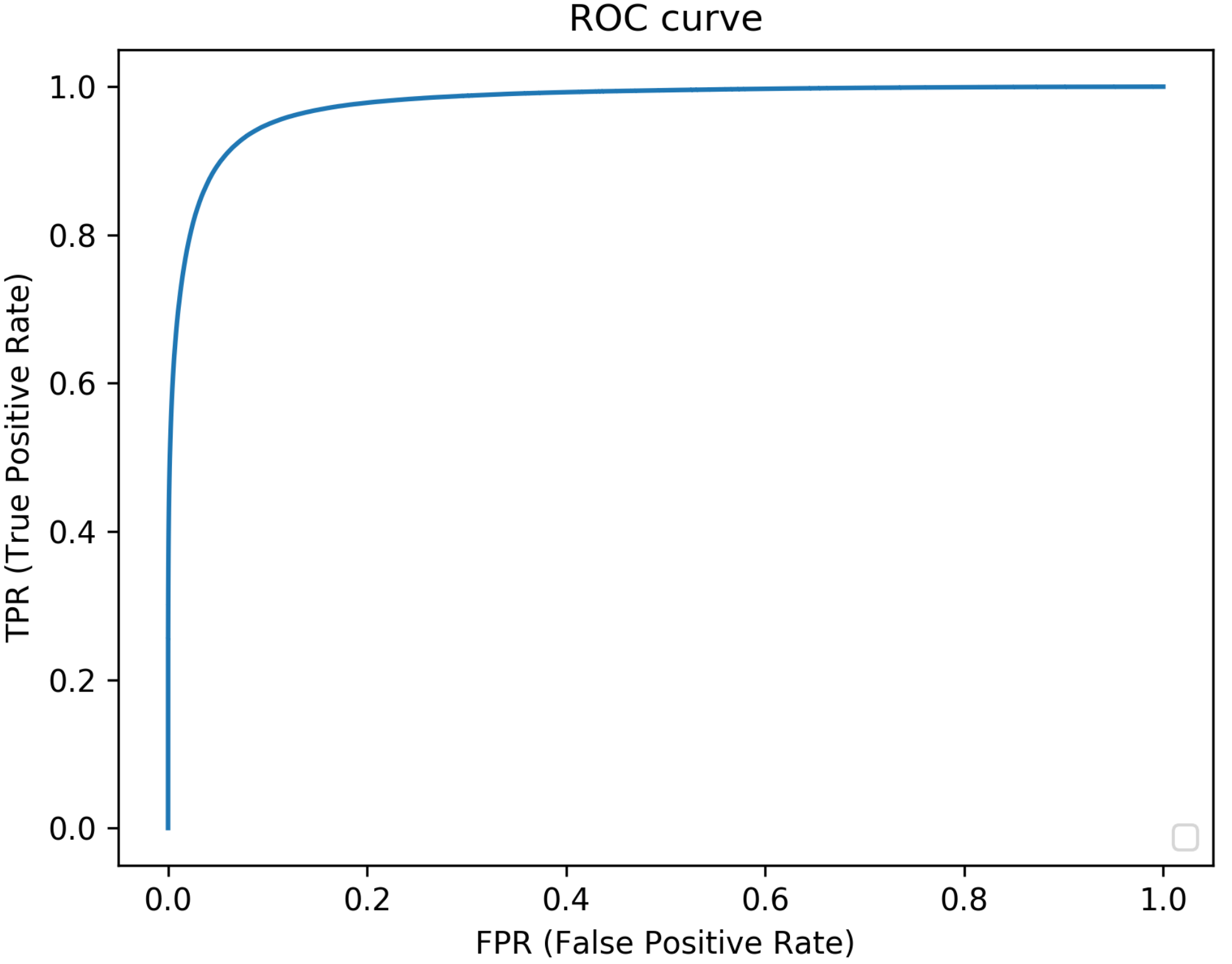}&
		\includegraphics[width=0.22\textwidth,height=25mm]{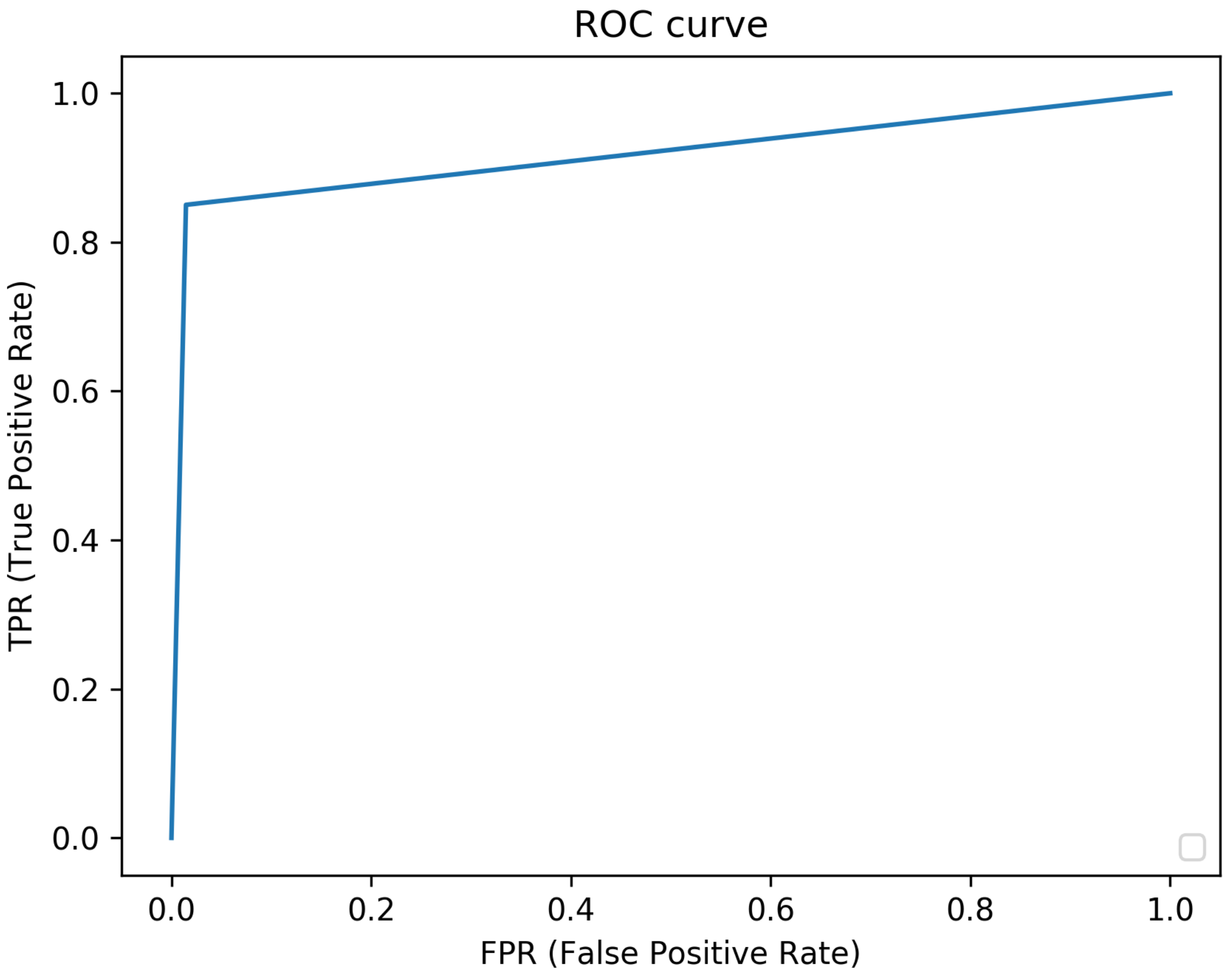}\\
		(a) DRIVE, & (b) ISIC 2017, \\
		\includegraphics[width=0.22\textwidth,height=25mm]{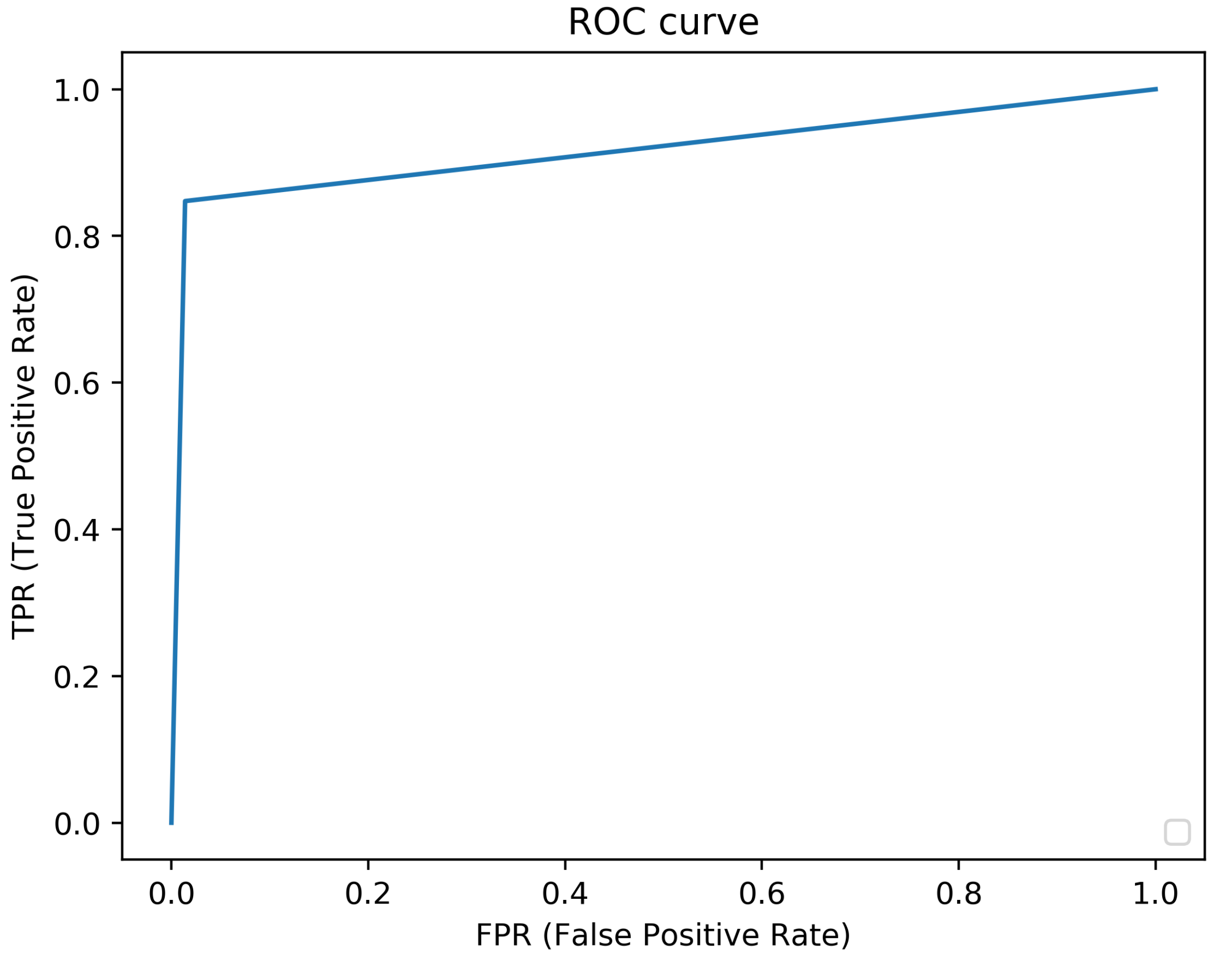} &
		\includegraphics[width=0.22\textwidth,height=25mm]{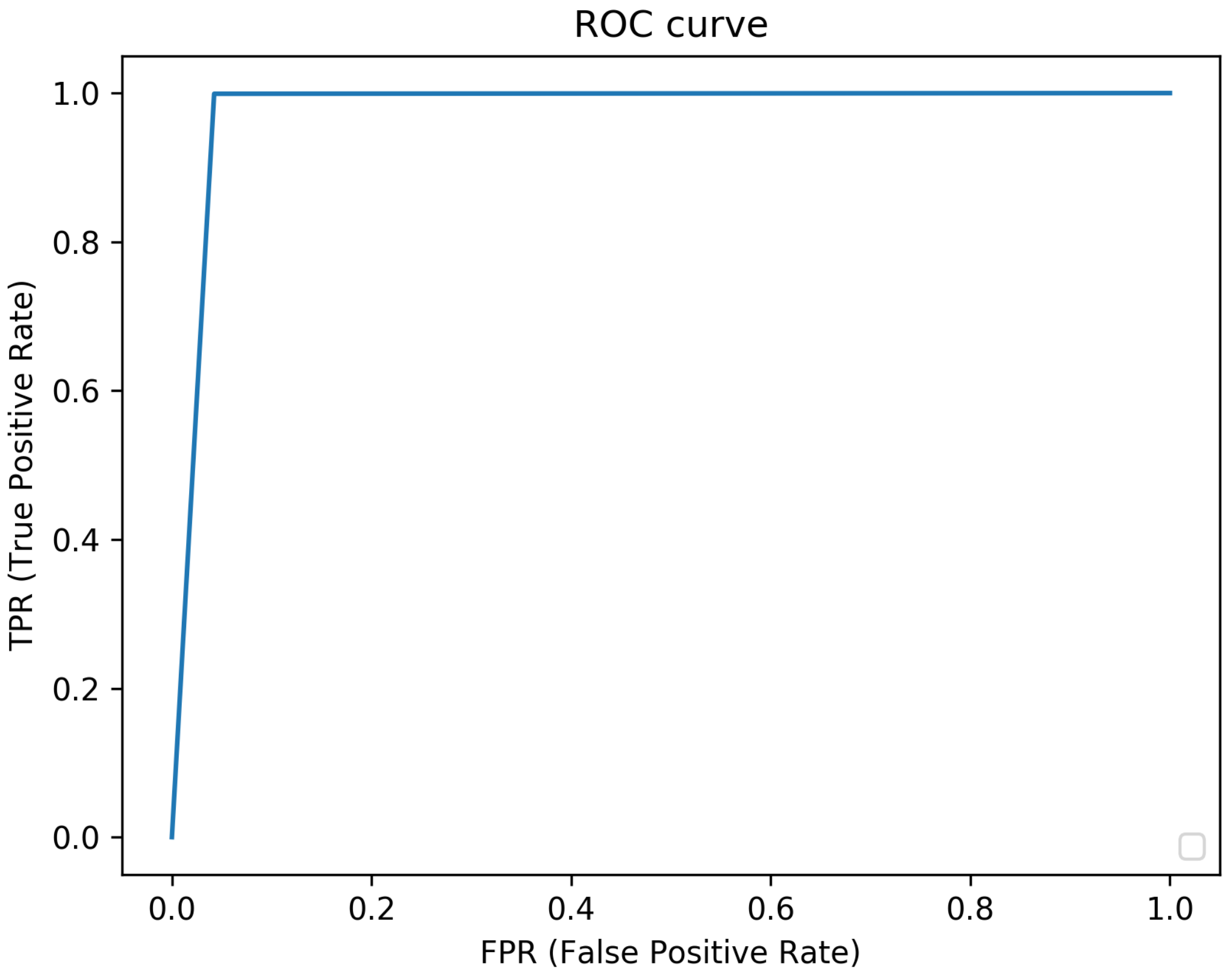}\\
		(c) ISIC 2018, & (d) Lung Segmentation,\\
		\includegraphics[width=0.22\textwidth,height=25mm]{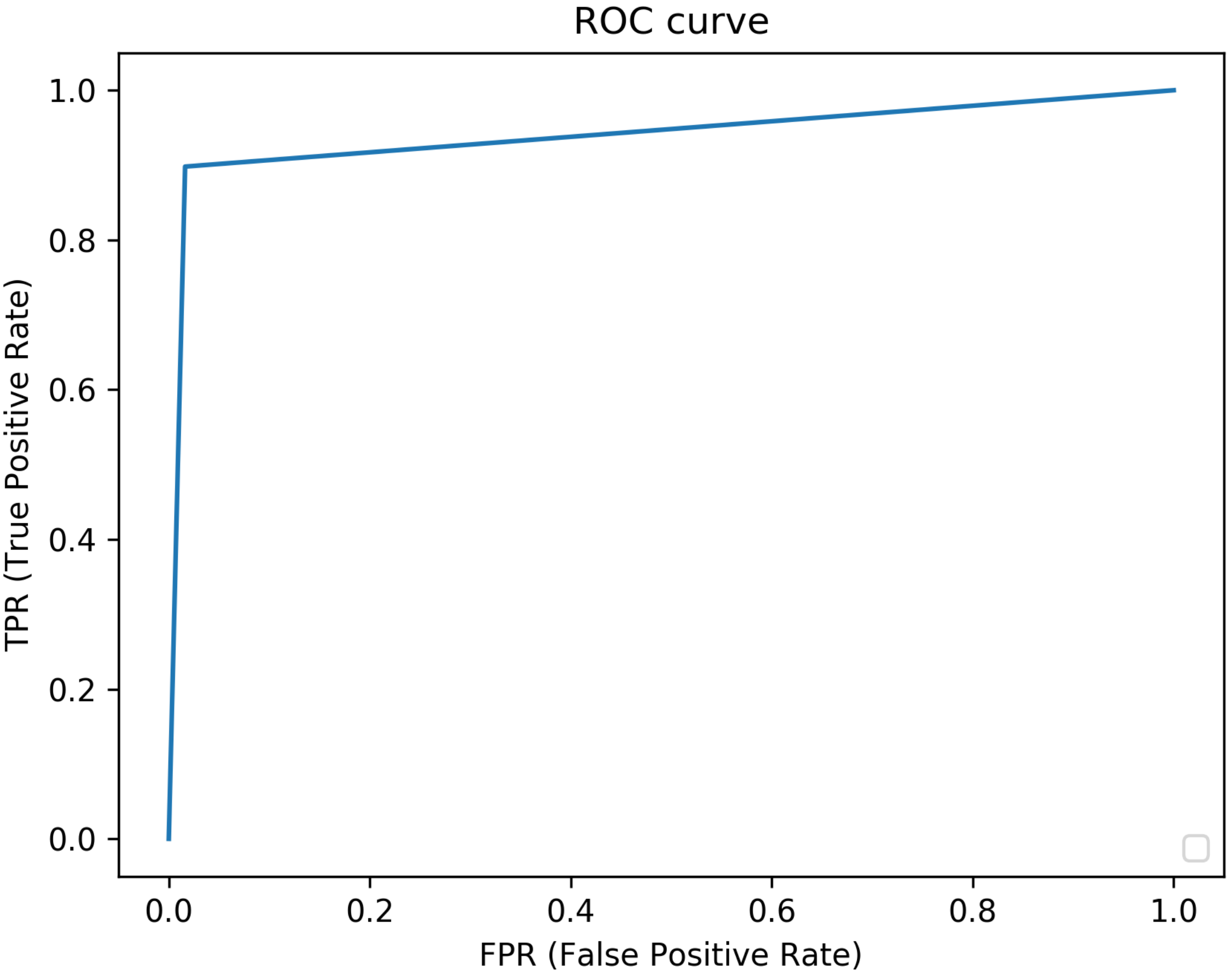}&
		\includegraphics[width=0.215\textwidth,height=25mm]{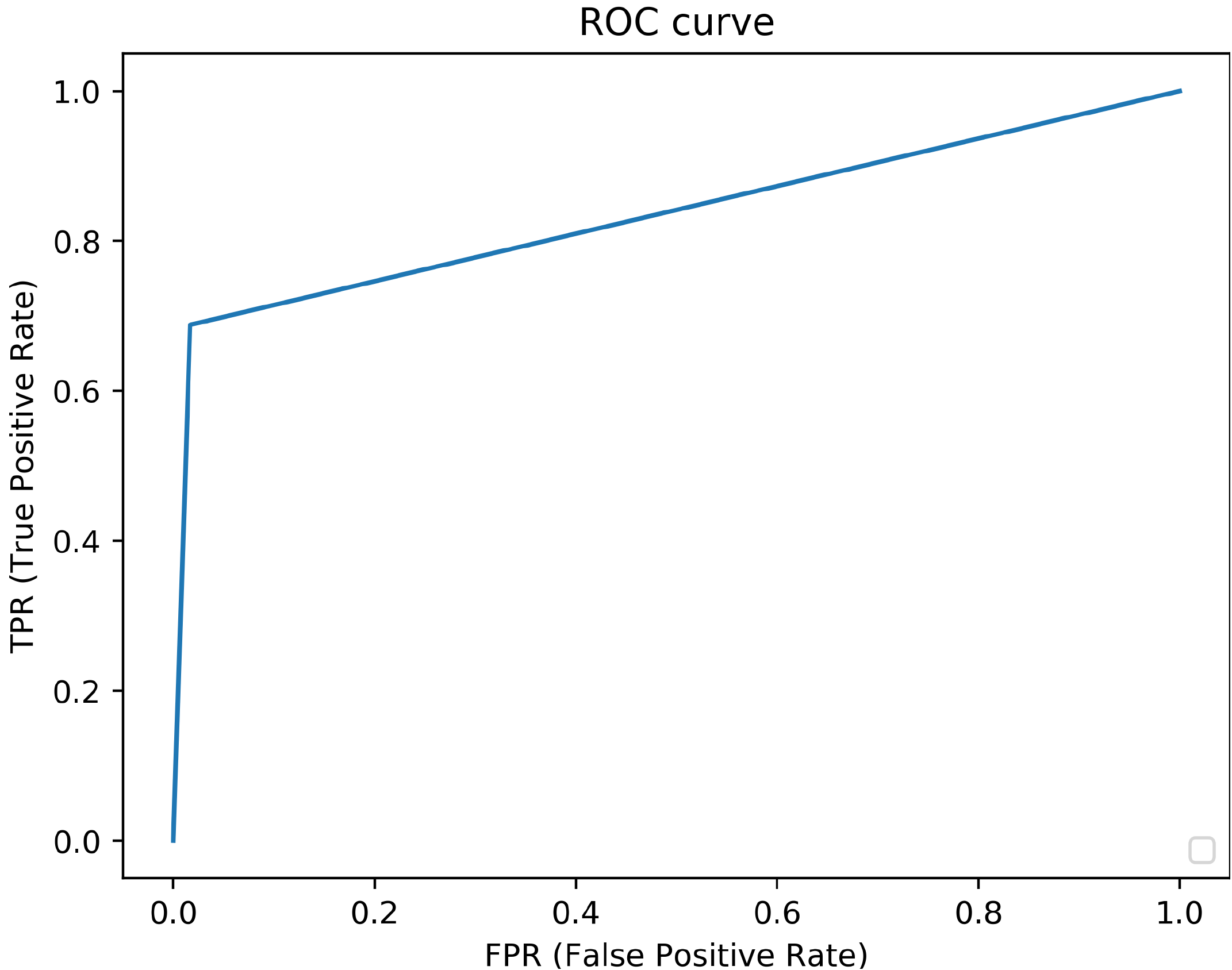} \\
		 (e) $PH^2$, & (f) Cell Nuclei Dataset\\
	\end{tabular}
	\caption{ROC diagrams of the proposed MCGU-Net for six dataset.}
	\vspace*{-\baselineskip}
	\label{fig:ROCs}
\end{figure}

To ensure the proper convergence of the proposed network, the training and validation accuracy for DRIVE dataset is shown in Figure \ref{fig:converge} (a). It is shown that the network converges very fast, i.e., after the $30^{th}$ epoch. %, the network is almost converged. 
We also can see that in the first 15 epochs the validation accuracy is larger than the training one. This fact is mostly because of the small size of dataset since we use a small set of images as the validation set. Moreover, it might be related to the fact that we evaluate the validation set at the end of epoch. To show the overall performance of the MCGU-Net on DRIVE dataset, ROC curves is shown in Figure \ref{fig:ROCs} (a). ROC is the plot of the true positive rate (TPR) against the false positive rate (FPR). %AUC (reported in Table \ref{tab:drive}) is the area under the ROC curve and is a measure of how well the network can segment the input data.

% \begin{figure*}[ht]
% 	\centering
% 	\begin{tabular}{ccc}
% 		% Requires \usepackage{graphicx}
% 		\includegraphics[width=0.24\textwidth,height=30mm]{Retina_ACC.pdf}&
% 		\includegraphics[width=0.24\textwidth,height=30mm]{ISIC17_ACC.pdf}&
% 		\includegraphics[width=0.24\textwidth,height=30mm]{ISIC18_ACC.pdf}\\
% 		(a) DRIVE, & (b) ISIC 2017, & (c) ISIC 2018, \\
% 		\includegraphics[width=0.24\textwidth,height=30mm]{Lung_ACC.pdf} &
% 		\includegraphics[width=0.24\textwidth,height=30mm]{Ph_ACC.pdf} &
% 		\includegraphics[width=0.24\textwidth,height=30mm]{Nuclei_ACC.pdf}  \\
% 		 (d) Lung Segmentation, & (e) $PH^2$,  & (f) Cell Nuclei Dataset\\
% 	\end{tabular}
% 	\caption{Training and validation accuracy of MCGU-Net for three datasets.}
% 	\vspace*{-\baselineskip}
% 	\label{fig:converge}
% \end{figure*}

% \begin{figure*}[ht]
% 	\centering
% 	\begin{tabular}{ccc}
% 		% Requires \usepackage{graphicx}
% 		\includegraphics[width=0.24\textwidth,height=30mm]{Retina_PR.pdf}&
% 		\includegraphics[width=0.24\textwidth,height=30mm]{ISIC17_PR.pdf}&
% 		\includegraphics[width=0.24\textwidth,height=30mm]{ISIC18_PR.pdf} \\
% 		(a) DRIVE, & (b) ISIC 2017, & (c) ISIC 2018, \\
% 		\includegraphics[width=0.24\textwidth,height=30mm]{Lung_PR.pdf}&
% 		\includegraphics[width=0.24\textwidth,height=30mm]{Ph_PR.pdf}&
% 		\includegraphics[width=0.24\textwidth,height=30mm]{Nuclei_PR.pdf} \\
% 		(d) Lung Segmentation, & (e) $PH^2$, & (f) Cell Nuclei Dataset\\
% 	\end{tabular}
% 	\caption{Precision-Recall diagrams of the proposed MCGU-Net for four dataset.}
% 	\vspace*{-\baselineskip}
% 	\label{fig:ROCs}
% \end{figure*}

% \begin{figure*}[ht]
% 	\centering
% 	\begin{tabular}{ccc}
% 		% Requires \usepackage{graphicx}
% 		\includegraphics[width=0.3\textwidth,height=30mm]{PC_DRIVE.pdf}&
% 		\includegraphics[width=0.3\textwidth,height=30mm]{PC_ISIC.pdf}&
% 		\includegraphics[width=0.3\textwidth,height=30mm]{PC_Lung.pdf}\\
% 		(a) DRIVE, & (b) ISIC, & (c) Lung Segmentation,\\
% 	\end{tabular}
% 	\caption{Precision-Recall diagrams of the proposed MCGU-Net for three dataset.}
% 	\label{fig:PCs}
% \end{figure*}
\subsection{ISIC 2017 Dataset}
The ISIC 2017 dataset \cite{codella2018skin} is taken from the Kaggle competition. % which consisted of 3 tasks of: lesion segmentation, dermoscopic feature detection, and disease classification that occurred in 2017. 
We evaluate the proposed method on the provided data for skin lesion segmentation. This dataset provides 2000 skin lesion images as a training set with masks (containing cancer or non-cancer lesions) for segmentation. We use 1250 samples for training, 150 samples as validation data, and the other 600 samples for test. The original size of each sample is $576\times767$. We %use the same pre-processing as \cite{alom2018} on the input image, and 
resize images to $256\times256$. For this dataset, we train the network with pre-trained weights on imageNet. Since the input data is RGB images, the pre-trained weights are good initialization for the network.

Figure \ref{fig:skin_R}(a) shows some segmentation results of the proposed network on ISIC 2017. In Table \ref{tab:isic17}, the results of the MCGU-Net on this dataset are compared with the state-of-the-art approaches. It can be seen that MCGU-Net with both $d=1$ and $d=3$ achieves better results (except sensitivity) than the other approaches. Moreover, the result of MCGU-Net with three dense blocks is a bit higher than with one dense block. 
The training and validation accuracy of the proposed network for this dataset is shown in Figure \ref{fig:converge} (b). The network converges very fast for this data (after the $30^{th}$ epoch). %The validation set contains somehow different images than training set, therefore, the validation accuracy is lower than the training data. 
To show the overall performance of the MCGU-Net on ISIC 2017 dataset, the ROC curves are shown in Figure \ref{fig:ROCs} (b).

\subsection{ISIC 2018 Dataset}
\begin{figure}[ht]
	\centering
	\begin{tabular}{ccc}
		\includegraphics[width=0.14\textwidth,height=25mm]{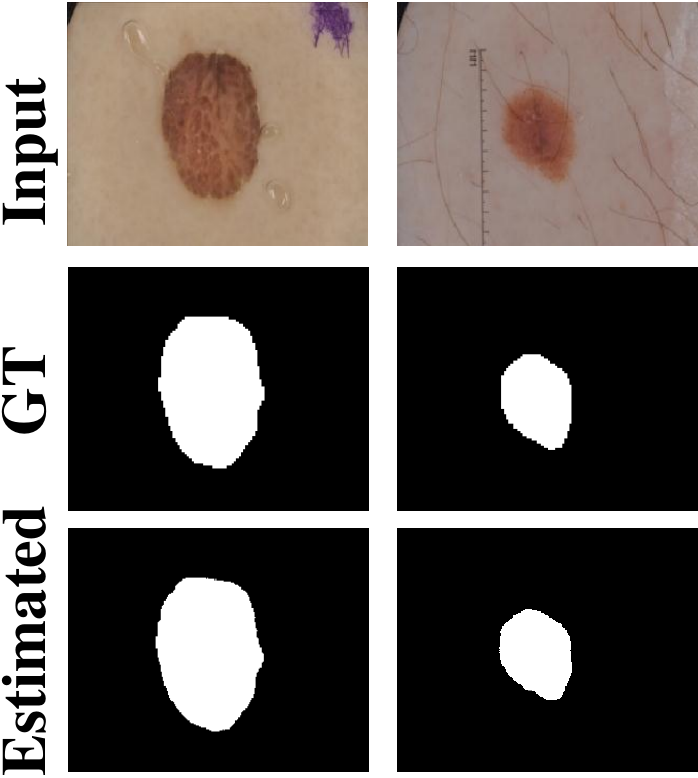}&
		\includegraphics[width=0.14\textwidth,height=25mm]{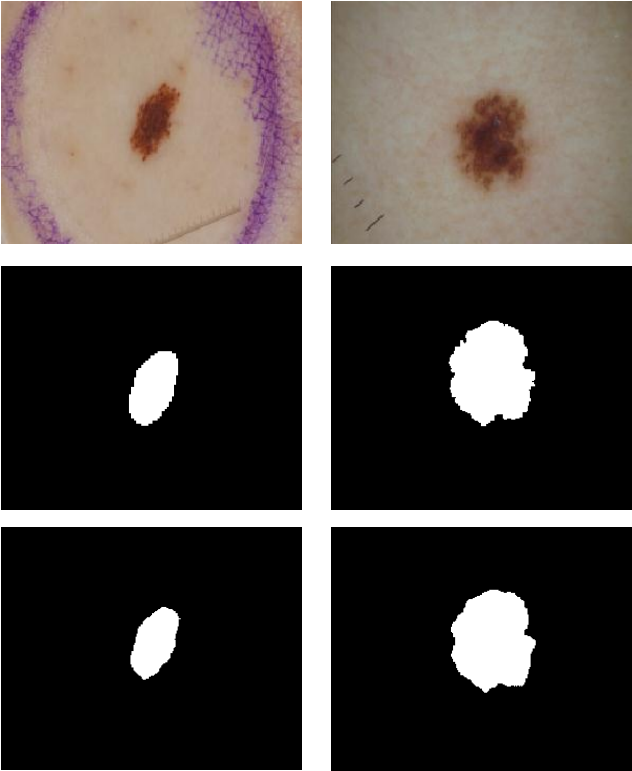}&
		\includegraphics[width=0.14\textwidth,height=25mm]{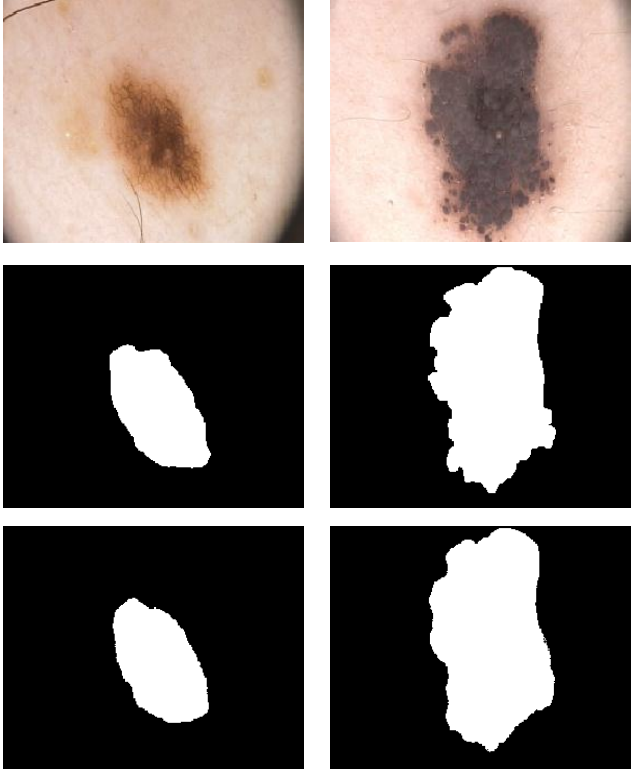}\\
		(a) ISIC 2017, & (b) ISIC 2018,& (c) $PH^2$
	\end{tabular}
	\caption{Segmentation result of MCGU-Net on three datasets.}
	\vspace*{-\baselineskip}
	\label{fig:skin_R}
\end{figure}

This dataset \cite{codella2019skin} %, shown in Figure \ref{fig:Datasets} (b), 
was published by the International Skin Imaging Collaboration (ISIC) as a large-scale dataset of dermoscopy images in 2018. %This dataset is taken from a challenge on lesion segmentation, dermoscopic feature detection, and disease classification. 
It includes 2594 images where like previous approaches\cite{alom2018}, we used 1815 images for training, 259 for validation and 520 for testing.
The original size of each sample is $2016\times3024$. We resize images to $256\times256$. The training data consists of the original images and corresponding ground truth annotations (containing cancer or non-cancer lesions). Like the ISIC 2017 dataset, the proposed network works better with pre-trained weights. 
For qualitative analysis, Figure \ref{fig:skin_R}(b) shows some example outputs of the proposed MCGU-Net on ISIC 2018. 
Table \ref{tab:isic18} lists the quantitative results obtained by different methods and the proposed network on this dataset. A large improvement is achieved by the MCGU-Net (with both $d=1$ and $d=3$) w.r.t. state-of-the-art alternatives for all of the evaluation metrics. It is clear that the network with $d=3$ works better than the one with $d=1$.
It is worth mentioning that there was a challenge on ISIC dataset and the best result achieved by the participants was $JS = 0.802$. Compare to this result, there is a good gap between the $JS$ achieved by the MCGU-Net ($0.955$) and the best result of the ISIC challenge.
%Compared to this result, the improvement of MCGU-Net (JS=$0.936$) if more than 0.13 of JS and the best result of the ISIC challenge.
%It is worth mentioning that the best result achieved by the participants in a challenge on ISIC dataset was JS$=0.802$.

The training and validation accuracy of the proposed network for ISIC dataset is shown in Figure \ref{fig:converge} (c). The convergence speed of the network for ISIC dataset is fast (after 40 epochs). The validation accuracy over the training process is variable. The reason behind this fact is that the validation set contains some images totally different from the ones in training set, therefore, during the first learning iterations the model has some problems about segmenting those images. To show the overall performance of the MCGU-Net on ISIC dataset, the ROC curves are shown in Figure \ref{fig:ROCs} (c).

%\begin{figure}
%\centering
%\includegraphics[width=0.3\textwidth]{ISIC17_Res2.pdf}
%\caption{Segmentation result of MCGU-Net on ISIC 2017.} 
%\vspace*{-\baselineskip}
%\label{fig:ISIC17_R}
%\end{figure}

%\begin{figure}
%\centering
%\includegraphics[width=0.45\textwidth]{Skin_Res.pdf}
%\caption{Segmentation result of MCGU-Net on ISIC 2017.} 
%\vspace*{-\baselineskip}
%\label{fig:skin_R}
%\end{figure}

%\begin{figure}
%\centering
%\includegraphics[width=0.45\textwidth]{Skin_Res.pdf}
%\caption{Segmentation result of MCGU-Net on ISIC 2017.} 
%\vspace*{-\baselineskip}
%\label{fig:skin_R}
%\end{figure}

\begin{table}
\centering
    %\vspace*{-\baselineskip}
	%\caption{Performance comparison of the proposed network and the state-of-the-art methods on ISIC 2017 dataset.}
	\caption{Performance comparison on ISIC 2017 dataset.}
	\begin{tabular}{cccccc}
		\hline
		\textbf{Methods} & \textbf{F1}&	\textbf{SE}&	\textbf{SP}&	\textbf{AC}&	\textbf{JS}\\
		\hline
		U-net  \cite{ronneberger2015} & 0.8682 & 0.9479&0.9263&	0.9314&		0.9314\\
		Melanoma det.  \cite{codella2018skin} & -&-&	-&	o.9340	& -\\
		Lesion Analysis  \cite{li2018skin} & - & 0.8250&	0.9750&	0.9340 &		- \\
		R2U-net  \cite{alom2018} & 0.8920 & \textbf{0.9414}&	0.9425 &	0.9424&		0.9421\\
		%BCDU-Net (d=1)& 0.847& 0.783 &  0.980&	0.936 &0.922 &0.936\\
		%BCDU-Net (d=3)&0.851& 0.785& 0.982&	 0.937& 0.928& 0.937\\
		%BCDU-Net \cite{azad2019bi}& 0.8763 & 0.8575 & 0.9747 & 0.9510 &  0.9510 \\
		%BCDU-Net \cite{azad2019bi}& 0.8797 & 0.8441 & 0.9810 & 0.9533 &  0.9533 \\
		%BCDU-Net \cite{azad2019bi}& 0.8810 & 0.8647 & 0.9751 & 0.9528 &  0.9528 \\
		\hline
		\textbf{MCGU-Net (d=1)}& 0.8871  & 0.8305  & \textbf{0.9888} & 0.9555  &  0.9555 \\
		\textbf{MCGU-Net (d=3)}& \textbf{ 0.8927}& \textbf{0.8502}& 0.9855&	\textbf{0.9570}&  \textbf{0.9570}\\
		\hline
	\end{tabular}
	\label{tab:isic17}
\end{table}

\begin{table}
\centering
    %\vspace*{-\baselineskip}
	%\caption{Performance comparison of the proposed network and the state-of-the-art methods on ISIC 2018 dataset.}
	\caption{Performance comparison on ISIC 2018 dataset.}
	\begin{tabular}{ccccccc}
		\hline
		\textbf{Methods} & \textbf{F1}&	\textbf{SE}&	\textbf{SP}&	\textbf{AC}&\textbf{PC}&	\textbf{JS}\\
		\hline
		U-net  \cite{ronneberger2015} & 0.647 & 0.708 &	0.964 & 0.890&	 0.779& 0.549\\
		Att U-net  \cite{oktay2018} & 0.665&	0.717 &	0.967&	 0.897 &	0.787&	 0.566\\
		R2U-net  \cite{alom2018} & 0.679 &	0.792 & 0.928& 0.880& 0.741& 0.581\\
		Att R2U-Net \cite{alom2018} &0.691&	0.726&	0.971&	0.904&	0.822&	 0.592\\
		%BCDU-Net (d=1)& 0.847& 0.783 &  0.980&	0.936 &0.922 &0.936\\
		%BCDU-Net (d=3)&0.851& 0.785& 0.982&	 0.937& 0.928& 0.937\\
		%BCDU-Net \cite{azad2019bi}&0.851& 0.785& 0.982&	 0.937& 0.928& 0.937\\
		\hline
		\textbf{MCGU-Net (d=1)}& 0.889 & 0.845 &  0.984 & 0.952 & 0.938 & 0.952\\
		\textbf{MCGU-Net (d=3)}& \textbf{0.895}& \textbf{0.848}& \textbf{0.986}&	\textbf{0.955}& \textbf{0.947}& \textbf{0.955}\\
		\hline
	\end{tabular}
	\label{tab:isic18}
\end{table}

\subsection{Lung Segmentation Dataset}
A lung segmentation dataset is introduced in the Lung Nodule Analysis (LUNA) competition at the Kaggle Data Science Bowl in 2017. This dataset %(shown in Figure \ref{fig:Datasets} (c)) 
consists of 2D and 3D CT images with respective label images for lung segmentation \cite{lungdata}. We use $70\%$ of the data as the train set and the remaining $30\%$ as the test set. The size of each image is $512\times 512$. For this dataset, the MCGU-Net works better with training from scratch since the input data is entirely different from images in ImageNet dataset.
%, however, we use a resized version of image ($256 \times 256$) as the input data to the proposed network. %For this dataset, we learn the surrounding part of lung tissue, instead of itself. To do that, we apply a pre-processing procedure (shown in Algorithm \ref{Alg:1}) over the input images.
Since the lung region in CT images have almost the same Hausdorff value with non-object of interests such as bone and air, it is worth to learn lung region by learning its surrounding tissues. To do that first we extract the surrounding region by applying algorithm \ref{Alg:1} and then make a new mask for the training sets. We train the model on these new masks and on the testing phase,and estimate the lung region as a region inside the estimated surrounding tissues. %A sample is shown in Figure \ref{fig:Alg1} .

\begin{figure}
	\centering
	\includegraphics[width=0.4\textwidth]{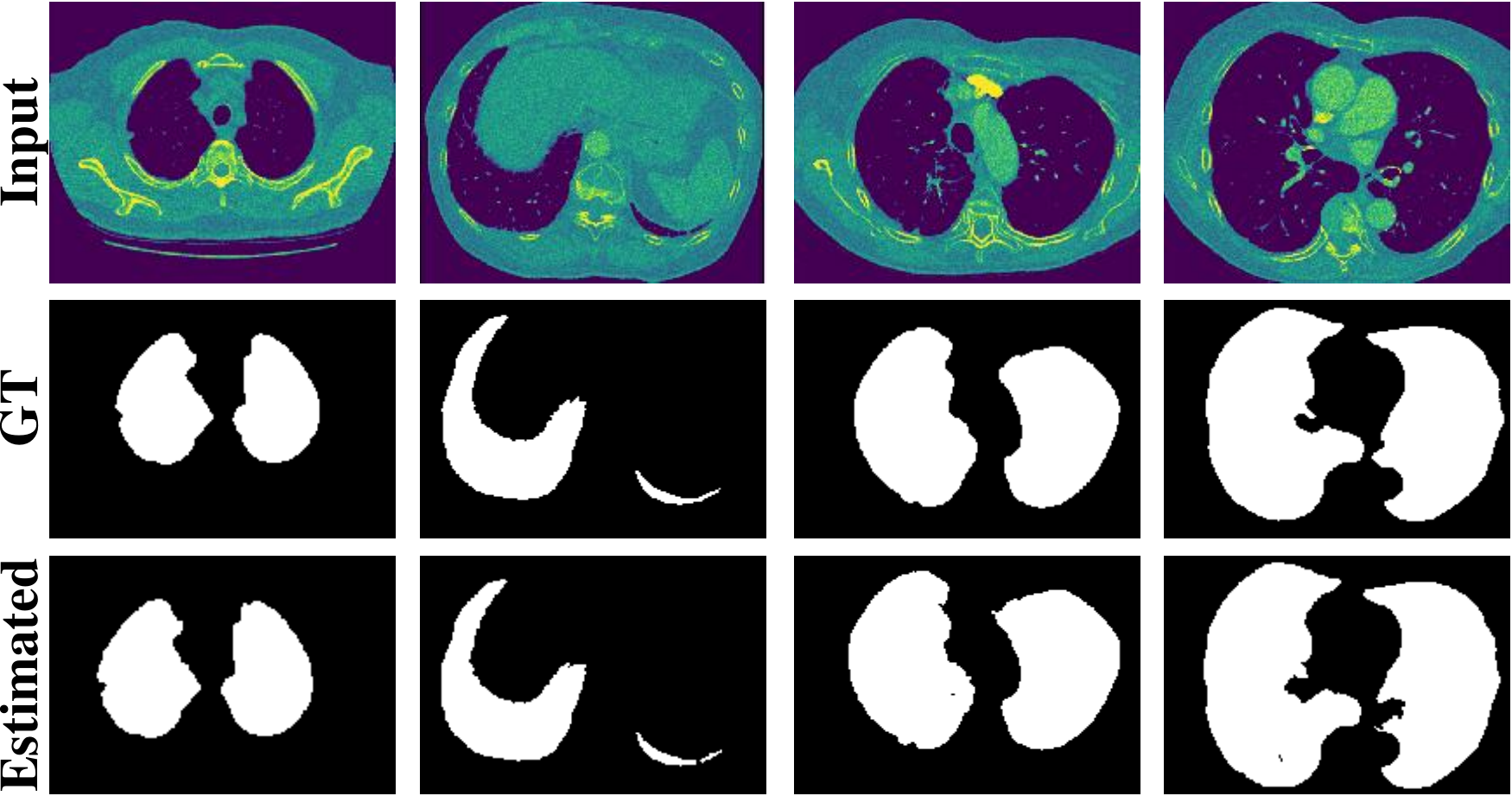}
	\caption{Segmentation result of MCGU-Net on Lung dataset.} 
	\vspace*{-0.5\baselineskip}
	\label{fig:Lung_R}
\end{figure}

Figure \ref{fig:Lung_R} shows some segmentation outputs of the proposed network for lung dataset. The quantitative results of the proposed MCGU-Net is compared with other methods in Table \ref{tab:lung}. It is clear that the MCGU-Net (with both $d=1$ and $d=3$) outperforms the other methods. Moreover, the network with dense connections works better. The training and validation accuracy for this dataset is shown in Figure \ref{fig:converge} (d). To show the overall performance of the network on this dataset, ROC curves is shown in Figure \ref{fig:ROCs} (d).

\begin{table}
\centering
    \vspace*{-\baselineskip}
	%\caption{Performance comparison of the proposed network and the state-of-the-art methods on Lung dataset.}
	\caption{Performance comparison on Lung dataset.}
	\begin{tabular}{cccccc}
		\hline
		\textbf{Methods} & \textbf{F1}&	\textbf{SE}&	\textbf{SP}&	\textbf{AC}&	\textbf{JS}\\
		\hline
		U-net  \cite{ronneberger2015} & 0.9658 & 0.9696&0.9872&	0.9872&		0.9858 \\
		RU-net  \cite{alom2018} &0.9638&	 0.9734&	0.9866&	0.9836&		0.9836 \\
		R2U-Net \cite{alom2018} &0.9832&	\textbf{0.9944}&	0.9832&	0.9918&		0.9918 \\
		%BCDU-Net (d=1)& 0.9889& 0.9901&0.9979&	0.9967&0.9940&0.9967\\
		%BCDU-Net (d=3)& 0.9904&0.9910& 0.9982&	 0.9972& 0.9946& 0.9972\\
		%BCDU-Net \cite{azad2019bi}& 0.9904&0.9910& 0.9982&	 0.9972& 0.9946& 0.9972\\
		\hline
		%\textbf{MCGU-Net (d=1)}& ?? & ?? & ?? &	?? & ?? & ??\\
		%\textbf{MCGU-Net (d=3)}& \textbf{0.9849}& 0.9852 &\textbf{ 0.9969} &	\textbf{ 0.9949} &\textbf{0.9911 } &\textbf{0.9949}\
		\textbf{MCGU-Net (d=1)}& 0.9889& 0.9901&0.9979&	0.9967&0.9967\\
		\textbf{MCGU-Net (d=3)}& \textbf{0.9904} & \textbf{0.9910}& \textbf{0.9982}& \textbf{0.9972}&  \textbf{0.9972}\\
		\hline
	\end{tabular}
	\label{tab:lung}
\end{table}

\begin{algorithm}
	\caption{Pre-processing over lung dataset.}
	\label{Alg:1}
	\begin{algorithmic}[1]
		\STATE Input = $X$ and $GT\_Mask$\\
		$Min\ range = -512$\\
		$Max\ range = +512$
		\STATE Output = $Surrounding\ Mask$
		\STATE $X(X>Max\ range) = Max\ range$ \\
		$X(X<Min\ range) = Min\ range$\\
		\COMMENT{Remove bones and vessels}
		\STATE $X = Norm(X)$
		\COMMENT{Normalize X}
		\STATE $X = image2binary(X)$
		\COMMENT{Convert to binary}
		\STATE $X = X \cup GT\_Mask$
		\STATE $X = Morphology (X)$
		\COMMENT{Remove noise}
		\STATE $Surrounding\_Mask = X-GT\_Mask$
	\end{algorithmic}
\end{algorithm}
%\vspace*{-\baselineskip}

\subsection{$PH^2$ Dataset}
 %The $PH^2$ \cite{mendoncca2013ph} was published through a joint research collaboration between the Universidade do Porto, Tecnico Lisboa, and the Dermatology service of Hospital Pedro Hispano in Matosinhos, Portugal. 
The $PH^2$ dataset \cite{mendoncca2013ph} is a a dermoscopic image database proposed for segmentation and classification. It contains a total number of 200 melanocytic lesions, including 80 common nevi, 80 atypical nevi, and 40 melanomas. The manual segmentations %and clinical diagnosis 
of the skin lesions are availablee. Each input image is a 8-bit RGB color images with a resolution of $768 \times 560$ pixels. There are not a pre-defined test and train sets for this dataset. %Therefore, for the fair comparison, 
We follow the experimental setting used in \cite{liu2019enhanced}. We randomly split the dataset into two sets of 100 images, and then use one set as the test data, $80\%$ of the other set for the train, and the remained data for the validation. Since the number of data is not enough for training the network, we exploit the learnt weights of ISIC 2017 as the pre-trained model (like \cite{liu2019enhanced}) and then finetune the network with train data. 

Some segmentation outputs of the proposed network for $PH^2$ dataset are depicted in Figure \ref{fig:skin_R}(c). In Table \ref{tab:ph}, the results of the proposed network are compared with other state-of-the-art approaches. We can see the MCGU-Net results in better performance than other methods. The network has almost the same performance for both $d=1$ and $d=3$. The reason behind this fact is the small size of training data since the network with $d=1$ contains fewer parameters for learning. The training and validation accuracy for this dataset is shown in Figure \ref{fig:converge} (e). The network converges very fast ($20^{th}$ epoch) which might be related to the small size of data. The %overall performance of the network on this dataset, 
ROC curve is shown in Figure \ref{fig:ROCs} (e).

%\begin{figure}
%\centering
%\includegraphics[width=0.3\textwidth]{Ph_Res.pdf}
%\caption{Segmentation result of MCGU-Net on $PH^2$ dataset.} 
%\vspace*{-\baselineskip}
%\label{fig:PH_R}
%\end{figure}

% \begin{table*}
%\centering
%    \vspace*{-\baselineskip}
%	\caption{Performance comparison of the proposed network and the state-of-the-art methods on $PH^2$ dataset.}
%	\begin{tabular}{cccccc}
%		\hline
%		\textbf{Methods} & \textbf{DIC}&	\textbf{Sensitivity}&	\textbf{Specificity}&	\textbf{Accuracy}&	\textbf{JS}\\
%		\hline
%		FCN \cite{noh2015learning} &  0.8903 & 0.9030 & 0.9402  & 0.9282 & 0.8022  \\
%		U-net  \cite{ronneberger2015} & 0.8761  & 0.8163 & 0.9776 &	0.9255 &	0.7795 \\
%		SegNet \cite{badrinarayanan2017segnet} & 0.8936 &	 0.8653 &	0.9661&	0.9336 &	0.8077  \\
%	    FrCN \cite{al2018skin} & 0.9177&	\textbf{0.9372}   &	0.9565&	0.9508 &	0.8479 \\
%		\hline
%		%\textbf{MCGU-Net (d=1)}& ?? & ?? & ?? &	?? & ?? & ??\\
%		%\textbf{MCGU-Net (d=3)}& \textbf{0.9849}& 0.9852 &\textbf{ 0.9969} &	\textbf{ 0.9949} &\textbf{0.9911 } &\textbf{0.9949}\
%		\textbf{MCGU-Net (d=1)}& 0.9762  &  \textbf{0.8727} & \textbf{0.9925} &	 0.9536 & 0.9536 \\
%		\textbf{MCGU-Net (d=3)}& \textbf{0.9763 } & 0.8322 & 0.9714 & \textbf{ 0.9537}& \textbf{0.9537 }\\
%		\hline
%	\end{tabular}
%	\label{tab:ph}
%\end{table*}

 \begin{table}
\centering
    \vspace*{-\baselineskip}
	%\caption{Performance comparison of the proposed network and the state-of-the-art methods on $PH^2$ dataset.}
	\caption{Performance comparison on $PH^2$ dataset.}
	\begin{tabular}{cccccc}
		\hline
		\textbf{Methods} & \textbf{DIC}&	\textbf{SE}&	\textbf{SP}&	\textbf{AC}&	\textbf{JS}\\
		\hline
		FCN \cite{noh2015learning} &  0.8903 & 0.9030 & 0.9402  & 0.9282 & 0.8022  \\
		U-net  \cite{ronneberger2015} & 0.8761  & 0.8163 & 0.9776 &	0.9255 &	0.7795 \\
		SegNet \cite{badrinarayanan2017segnet} & 0.8936 &	 0.8653 &	0.9661&	0.9336 &	0.8077  \\
	    FrCN \cite{al2018skin} & 0.9177&	\textbf{0.9372}   &	0.9565&	0.9508 &	0.8479 \\
		\hline
		%\textbf{MCGU-Net (d=1)}& ?? & ?? & ?? &	?? & ?? & ??\\
		%\textbf{MCGU-Net (d=3)}& \textbf{0.9849}& 0.9852 &\textbf{ 0.9969} &	\textbf{ 0.9949} &\textbf{0.9911 } &\textbf{0.9949}\
		\textbf{MCGU-Net (d=1)}& 0.9762  &  \textbf{0.8727} & \textbf{0.9925} &	 0.9536 & 0.9536 \\
		\textbf{MCGU-Net (d=3)}& \textbf{0.9763 } & 0.8322 & 0.9714 & \textbf{ 0.9537}& \textbf{0.9537 }\\
		\hline
	\end{tabular}
	\label{tab:ph}
\end{table}

\subsection{Cell Nuclei Dataset}
\begin{figure}
 \centering
 \includegraphics[width=0.4\textwidth]{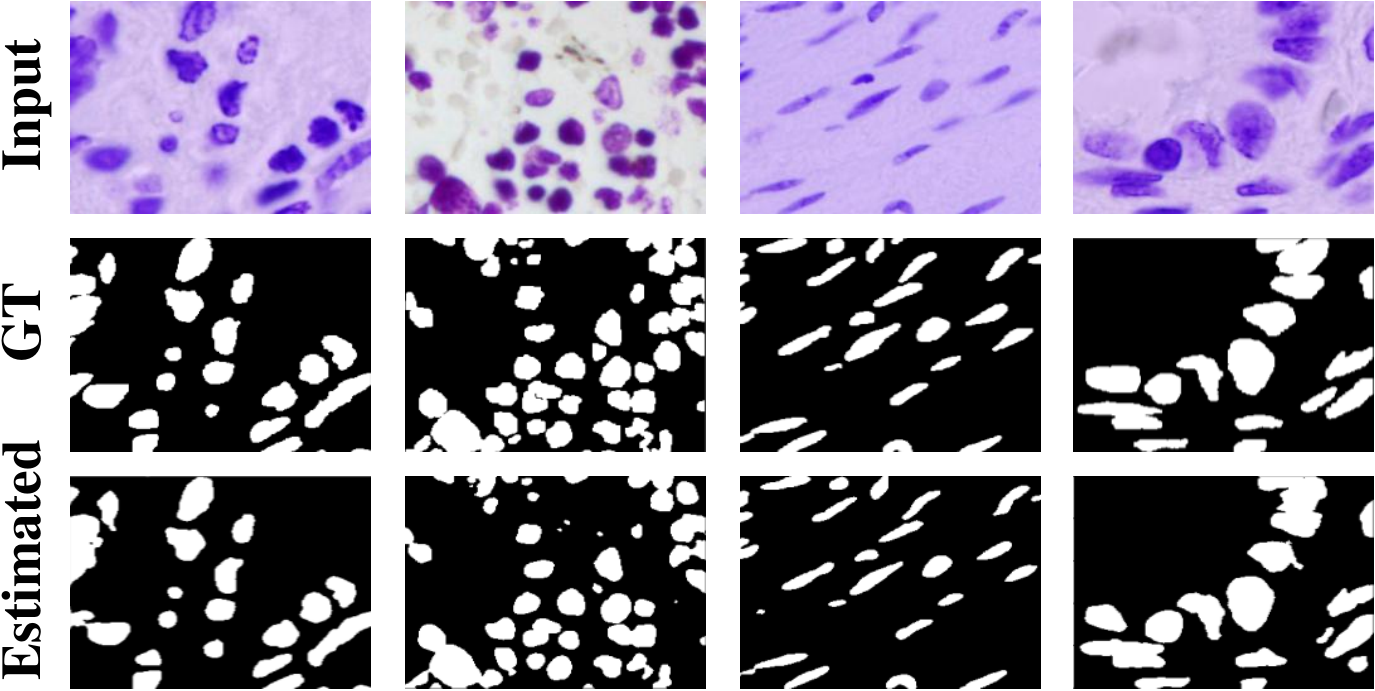}
 \caption{Segmentation result of MCGU-Net on cell nuclei Dataset.} 
 \vspace*{-0.5\baselineskip}
 \label{fig:Nuclei_R}
 \end{figure}

We evaluate the proposed network on the dataset from 2018 Kaggle Data Science Bowl 2018 \cite{bowl2018}. This data is captured with various situations, like different cell type, illumination status, and image size. Moreover, this dataset contains smaller regions inside images for segmentation for which we want to evaluate the performance of the MCGU-Net. It includes a total number of 670 images. We randomly split the data into $70\%$ training, $10\%$ validation, and $20\%$ test data sets.
Figure \ref{fig:Nuclei_R} shows some segmentation outputs of MCGU-Net. In Table \ref{tab:Nuclei}, the performance of the proposed method is compared with other approaches. We can see there is a high gap between the results of the MCGU-Net and other methods. The network works better with $d=3$ for this data. The training and validation accuracy for this dataset is shown in Figure \ref{fig:converge} (f). Since the validation and training data are taken from the same set (and validation set is smaller than train set), the validation accuracy is a bit higher than the training. The %overall performance of the network on this dataset,
 ROC curve is shown in Figure \ref{fig:ROCs} (f).

%\begin{table*}
%\centering
%    \vspace*{-\baselineskip}
%	\caption{Performance comparison of the proposed network and the state-of-the-art methods on Cell Nuclei dataset.}
%	\begin{tabular}{cccccc}
%		\hline
%		\textbf{Methods} & \textbf{F1}&	\textbf{DIC}&	\textbf{Accuracy}&	\textbf{JS}\\
%		\hline
%		U-net  \cite{ronneberger2015}  &  0.8994 & 0.9077 & 0.9604  & 0.8310  \\
%	    Attention U-Net \cite{oktay2018} & 0.8899    & 0.8879 & 	0.9672 &	0.7984 \\
%		FocusNet \cite{kaul2019focusnet} & 0.8998 &	 0.8996 &	0.9697 &	0.8176   \\
%	    \hline
%		\textbf{MCGU-Net (d=1)}& 0.9295  &  0.9882 & 0.9766 &	 0.9766\\
%		\textbf{MCGU-Net (d=3)}& \textbf{0.9306} & \textbf{0.9884}& \textbf{0.9771 }& \textbf{ 0.9771}\\
%		\hline
%	\end{tabular}
%	\label{tab:Nuclei}
%\end{table*}

\begin{table}
\centering
    \vspace*{-\baselineskip}
	%\caption{Performance comparison of the proposed network and the state-of-the-art methods on Cell Nuclei dataset.}
	\caption{Performance comparison on Cell Nuclei dataset.}
	\begin{tabular}{cccccc}
		\hline
		\textbf{Methods} & \textbf{F1}&	\textbf{DIC}&	\textbf{AC}&	\textbf{JS}\\
		\hline
		U-net  \cite{ronneberger2015}  &  0.8994 & 0.9077 & 0.9604  & 0.8310  \\
	    Att U-Net \cite{oktay2018} & 0.8899    & 0.8879 & 	0.9672 &	0.7984 \\
		FocusNet \cite{kaul2019focusnet} & 0.8998 &	 0.8996 &	0.9697 &	0.8176   \\
	    \hline
		\textbf{MCGU-Net (d=1)}& 0.9295  &  0.9882 & 0.9766 &	 0.9766\\
		\textbf{MCGU-Net (d=3)}& \textbf{0.9306} & \textbf{0.9884}& \textbf{0.9771 }& \textbf{ 0.9771}\\
		\hline
	\end{tabular}
	\label{tab:Nuclei}
\end{table}

\subsection{Discussion}
The proposed network has some modifications from the original U-Net. We evaluate each modified part of the network to analyze its influence on the result.

%\subsubsection{Batch Normalization}
We included BN after each up-convolutional layer to speed up the network learning process. To evaluate the effect of this function, we train the network with and without BN. %Figure \ref{fig:BN} shows the training and validation accuracy of MCGU-Net for ISIC 2018 dataset without and with BN. %and Figure \ref{fig:BN} (b) shows the same contend for the network with BN.
 %MCGU-Net converged after 200 epochs without BN while this number is about 30 with BN, i.e., 
 %We can see that 
 BN yields the network to converge $2$ times faster. %Moreover, it can be seen that BN has improved the accuracy of the MCGU-Net. The variations among data in the ISIC dataset is larger than the other datasets. 
 BN manages the variations among data by standardizing data through controlling the mean and variance of distributions of inputs which results in a small regularization and reducing generalization error. %Therefore, BN helps the network to improve the performance.

%\begin{figure}[ht]
%	\centering
%	%\vspace{-5mm}
%	\begin{tabular}{cc}
%		% Requires \usepackage{graphicx}
%		\includegraphics[width=0.21\textwidth,height=25mm]{without_BN.pdf}&
%		\includegraphics[width=0.21\textwidth,height=25mm]{with_BN.pdf}\\
%		(a) & (b)\\
%	\end{tabular}
%	%\vspace{-5mm}
%	\caption{Training and validation accuracy of MCGU-Net (a) without and (b) with BN.}
%	\vspace*{-\baselineskip}
%	\label{fig:BN}
%	%\vspace{-6mm}
%\end{figure}

The last convolutional layer of the encoding path is augmented with dense blocks. The results for the network with $1$ and $3$ dense blocks are reported for all datasets. In Tables \ref{tab:drive} to \ref{tab:Nuclei}, it can be seen that MCGU-Net with $3$ dense block results in better performance.
The key idea of dense convolutions is collecting knowledge by sharing feature maps between blocks through direct connection between convolutional block. Consequently, each dense block receives all preceding layers as input, and therefore, produces more diversified and richer features. Thus, it helps the network to increase the representational power of deeper models.
We have more feature propagation both in backward and forward paths through dense blocks. The network performs a kind of deep supervision in backward path since dense block receives additional supervision from loss function through shorter connections \cite{huang2017densely}. The error signal is propagated to earlier layers more directly, hence, earlier layers can get direct supervision from the final softmax layer, and moreover, it results in decreasing the vanishing-gradient problem. 
In addition, compared to other deep architectures like residual connections, dense convolutions require fewer parameters while improving the accuracy of the network.
%\subsubsection{Dense Layer}
%Table \ref{tab:sum} shows that the network with dense connections improve the accuracy and F1-Score for the three datasets.

\begin{figure}
%\vspace{-5mm}
	\centering
	\includegraphics[width=0.4\textwidth]{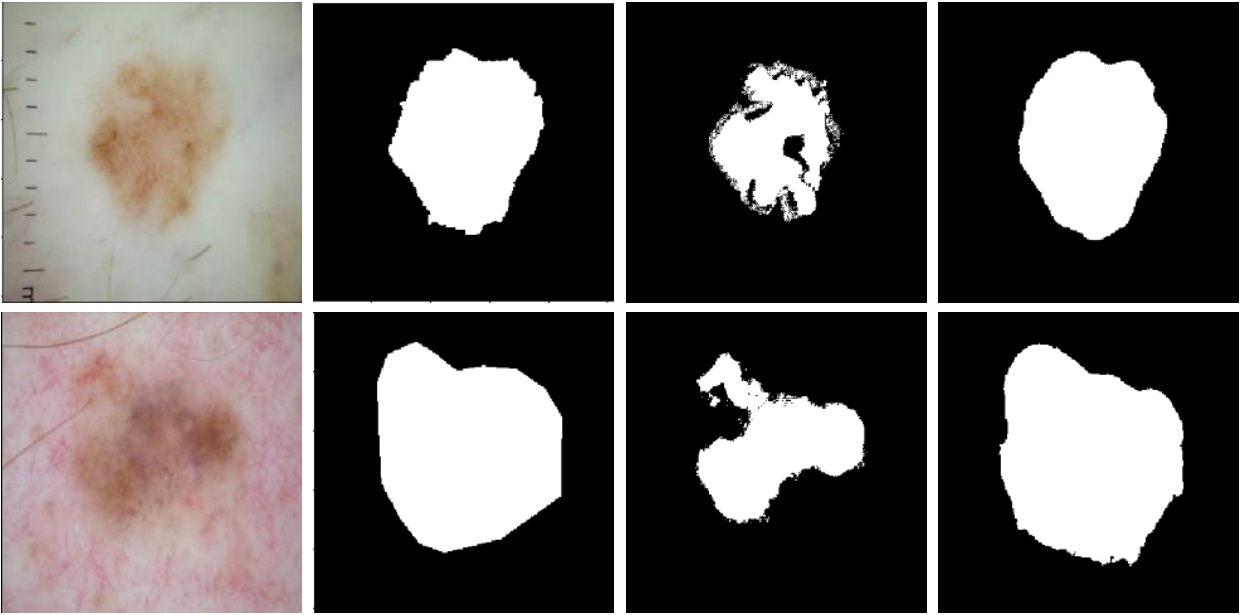}
	\caption{Visual effect of BConvLSTM in MCGU-Net . From left column 1 input, 2 GT mask, 3 and 4 are the outputs of network without and with ConvLSTM.} 
	 \vspace*{-\baselineskip}
	\label{figConvLSTM}
\end{figure}

In the proposed network, we used multi-level BConvLSTMs to combine encoded and decoded features. The encoded features have higher resolution and therefore contain more local information of the input image, while the decoded features have more semantic information about the input images. The affection of these two features over each other might result in a set of feature maps rich in both local and semantic information. Therefore, instead of a simple concatenation, we utilize BConvLSTM to combine the encoded and decoded features. In BConvLSTM, a set of convolution filters are applied on each kind of features. Therefore each ConvLSTM state, corresponds to one kind of features (e.g. encoded features), ia able to encode relevant information about the other kind of features (e.g. decoded features). %Finally, a non-linear hyperbolic tangent function is employed to combine the output of each ConvLSTM states. 
The convolutional filters along with the hyperbolic tangent functions help the network to learn complex data structures.
Figure \ref{figConvLSTM} shows the output segmentation mask of the original U-Net and MCGU-Net for two samples of the ISIC 2018 dataset. It shows a more precise and fine segmentation output of the proposed network. % than the original U-Net. 
%In each layer of the expanding path of the original U-Net (and different extensions of this network), 
%After the skip connections, there are two kinds of features to combine, the features from the previous decoding layer and the features from the corresponding encoding layer. For convenience, we call them the encoded and decoded features. In the original U-Net, a simple concatenation function is used to combine these two kinds of features.

%We summarized the "Accuracy" and "F1-Score" of the original U-Net and its modifications for three utilized datasets in Table \ref{tab:sum}. We evaluate each modified part of the network to analyze its influence on the result. In Table \ref{tab:sum}, it can be seen that the result of the standard U-Net is improved by inserting BConvLSTM in the skip connections. 

\begin{figure}
%\vspace{-5mm}
	\centering
	\includegraphics[width=0.4\textwidth]{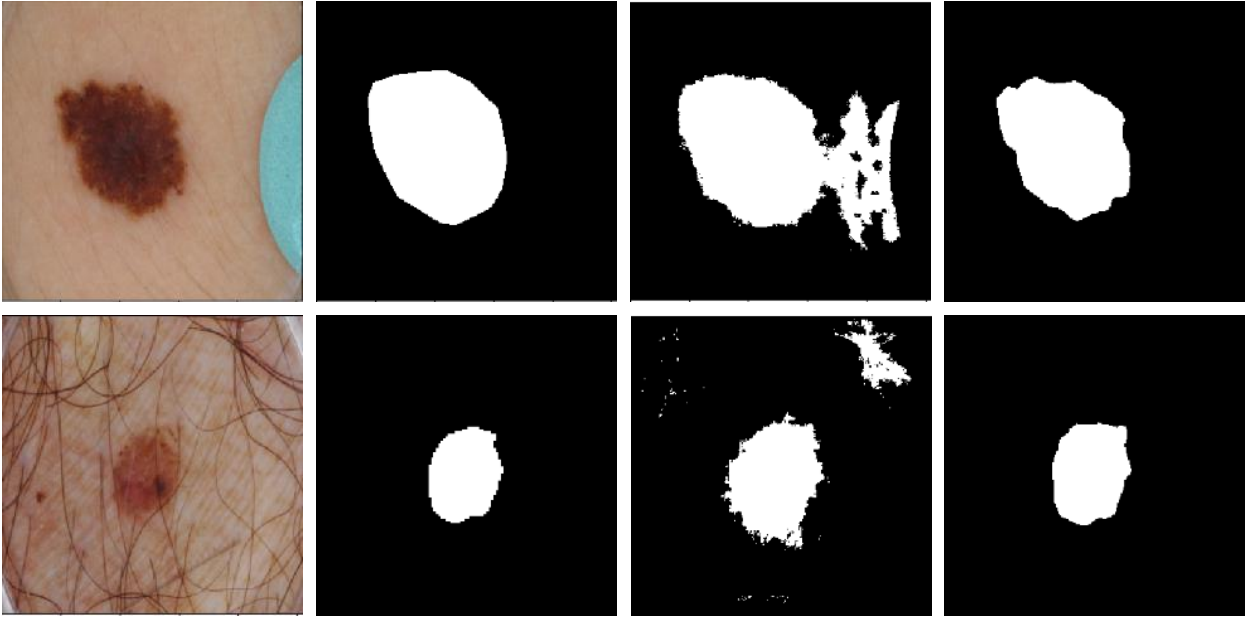}
	\caption{Visual effect of SE blocks in MCGU-Net.  From left column 1 input, 2 GT mask, 3 and 4 are the outputs of network without and with SE blocks. } 
	 \vspace*{-\baselineskip}
	\label{figSE}
\end{figure}

In Figure \ref{figSE}, we compare the segmentation output of the MCGU-Net with and without SE blocks for two samples of ISIC 2018 dataset, which demonstrates the power of SE features on semantic segmentation. It can be seen that the SE blocks help the network to produce more precise output by a context gating mechanism. To do that, this block exploits the global information embedded features in different channels and assign different channel attentions. The quality of the segmentation output of a network relies on effective feature learning. These findings reveal that the adaptive feature recalibration of SE blocks result in boosting the representational power of deep networks by focusing on informative features and suppressing weak ones. 

% \begin{table*}
% \centering
% 	\caption{Performance comparison of U-Net and its modifications in our work.}\label{tab2}
% 	\begin{tabular}{@{\extracolsep{4pt}}ccccccc@{}}
% 		\hline
% 		\multirow{2}{*}{\textbf{\textbf{Methods}}}  &\multicolumn{2}{c}{\textbf{DRIVE}} &	\multicolumn{2}{c}{\textbf{ISIC}} &\multicolumn{2}{c}{\textbf{Lung}}\\
% 		\cline{2-3} \cline{4-5} \cline{6-7}
% 		& \textbf{F1-Score} & \textbf{AC} & \textbf{F1-Score} & \textbf{AC} & \textbf{F1-Score} & \textbf{AC} \\
% 		\hline
% 		U-net  &0.8142 &0.9531 &0.6470 &0.8900 & 0.9658& 0.9828\\
% 		U-Net + BConvLSTM (d=1) &0.8222 &0.9559 &0.8470 &0.9360 &0.9889 &0.9967 \\
% 		U-Net + BConvLSTM + Dense Conv (d=3)  &0.8243 &0.9560 &0.8506 & 0.9374& 0.9904& 0.9972\\
% 		\hline
% 	\end{tabular}
% 	\label{tab:sum}
% \end{table*}
 %\vspace*{-\baselineskip}

%It can be seen that the result of the standard U-Net is improved by inserting BConvLSTM in the skip connections. Moreover, dense convolution can aff
%\subsubsection{BConvLSTM}

\section{Conclusion}

We proposed MCGU-Net for medical image segmentation. We showed that by including multi-level BConvLSTM in the skip connection, SE blocks in decoding path, inserting a densely connected convolutional blocks, and also employing SE blocks in decoding path, the network is able to capture more discriminative information which resulted in more precise segmentation results. Moreover, we were able to speed up the network by utilizing BN after the up-convolutional layer. The experimental results on six public benchmark datasets showed high gain in semantic segmentation in relation to state-of-the-art alternatives. %\footnote{Code and trained models will be released in a public github page after publication of the paper.}

\end{document}